\documentclass[modern]{aastex61}
\usepackage{natbib}
\usepackage{float}
\usepackage{graphicx}

\newcommand{\documentname}{\textsl{Article}}
\newcommand{\sectionname}{Section}
\newcommand{\equationname}{equation}
\newcommand{\project}[1]{\textsl{#1}}
\newcommand{\foreign}[1]{\textsl{#1}}
\newcommand{\code}[1]{\texttt{\detokenize{#1}}}

\newcommand{\etal}{\foreign{et al.}}

\newcommand{\acronym}[1]{{\scalebox{0.9}{#1}}}

\newcommand{\notename}{footnote}
\newcommand{\note}[1]{\footnote{#1}}

\newcommand{\problemname}{Problem}
\newcounter{problem}
\newenvironment{problem}{\refstepcounter{problem}\par\medskip\noindent\textbf{\problemname~\theproblem:}}{}

\newcommand{\setof}[1]{\{{#1}\}}
\newcommand{\dd}{\mathrm{d}}
\newcommand{\given}{\,|\,}
\newcommand{\mean}[1]{\left<{#1}\right>}

\usepackage{color,hyperref}
\definecolor{linkcolor}{rgb}{0,0,0.5}
\hypersetup{colorlinks=true,linkcolor=linkcolor,citecolor=linkcolor,
            filecolor=linkcolor,urlcolor=linkcolor}
\newcommand{\arxiv}[1]{\href{http://arxiv.org/abs/#1}{\textsl{arXiv}:#1}}
\newcommand{\isbn}[1]{\textsc{isbn:}#1}

\usepackage{needspace}

\shorttitle{using markov chain monte carlo}
\shortauthors{hogg \& foreman-mackey}

\newcommand{\MCMC}{\acronym{MCMC}}
\newcommand{\MH}{\acronym{M--H}}

\newcommand{\data}{D}
\newcommand{\pars}{\theta}
\newcommand{\best}{{\mathrm{(best)}}}
\newcommand{\better}{{\mathrm{(better)}}}

\begin{document}\sloppy\sloppypar\raggedbottom\frenchspacing\thispagestyle{plain}%
\title{Data analysis recipes:\\
       Using Markov Chain Monte Carlo%
\note{\label{note:first}%
Copyright 2013, 2014, 2015, 2016, 2017 the authors (OMG this took us a long time).
        This work is licensed under a
        \href{http://creativecommons.org/licenses/by-nc-nd/3.0/deed.en\_US}{%
            Creative Commons Attribution-NonCommercial-NoDerivs 3.0 Unported
            License}.}}%

\author[0000-0003-2866-9403]{David W. Hogg}
\affil{Center for Computational Astrophysics, Flatiron Institute, 162 Fifth Ave, New York, NY 10010, USA}
\affil{Center for Cosmology and Particle Physics, Department of Physics, New York University, 726 Broadway, New York, NY 10003, USA}
\affil{Center for Data Science, New York University, 60 Fifth Ave, New York, NY 10011, USA}
\affil{Max-Planck-Institut f\"ur Astronomie, K\"onigstuhl 17, D-69117 Heidelberg}

\author[0000-0002-9328-5652]{Daniel Foreman-Mackey}
\affil{Center for Computational Astrophysics, Flatiron Institute, 162 Fifth Ave, New York, NY 10010, USA}
\affil{NASA Sagan Fellow; Department~of~Astronomy, University~of~Washington, Box 351580, Seattle, WA 98195, USA}

\begin{abstract}\noindent
Markov Chain Monte Carlo (\MCMC) methods for sampling probability
density functions (combined with abundant computational resources)
have transformed the sciences, especially in performing probabilistic
inferences, or fitting models to data.
In this primarily pedagogical contribution,
we give a brief overview of the most basic \MCMC\ method and some
practical advice for the use of \MCMC\ in real inference problems.
We give advice on method choice, tuning for performance,
methods for initialization, tests of convergence, troubleshooting, and use of the chain
output to produce or report parameter estimates with associated uncertainties.
We argue that autocorrelation time is the most important test for convergence,
as it directly connects to the uncertainty on the sampling estimate of any
quantity of interest.
We emphasize that sampling is a method for doing integrals; this guides our thinking
about how \MCMC\ output is best used.
\end{abstract}

\section{When do you need MCMC?}\label{sec:when}

Markov Chain Monte Carlo (\MCMC) methods are methods for sampling
probability distribution functions or probability density functions (pdfs).
These pdfs may be either probability mass functions on a discrete
space, or probability densities on a continuous space, though we will
concentrate on the latter in this \documentname.
\MCMC\ methods don't require that you have a full analytic description of the
properly normalized pdf for sampling to proceed; they only require
that you be able to compute ratios of the pdf at pairs of locations.
This makes \MCMC\ methods ideal for sampling \emph{posterior
  pdfs} in probabilistic inferences:

In a probabilistic inference, the posterior pdf $p(\pars\given\data)$,
or pdf for the parameters $\pars$ given the data $\data$, is
constructed from the likelihood $p(\data\given\pars)$, or pdf for the
data given the parameters, and the prior pdf $p(\pars)$ for the
parameters by what's often known as ``Bayes rule'',
\begin{eqnarray}
p(\pars\given\data) &=& \frac{1}{Z}\,p(\data\given\pars)\,p(\pars)
\label{eq:bayes}\quad .
\end{eqnarray}
In these contexts, the constant $Z$, sometimes written as
$p(\data)$, is known by the names ``evidence'', ``marginal likelihood'',
``Bayes integral'', and ``prior predictive probability'', and is usually
\emph{extremely hard to calculate}.\note{%
  The factor $Z$ is often difficult to compute, because the
  likelihood (or the prior) can have extremely complex structure,
  with multiple arbitrarily compact modes, arbitrarily positioned
  in the (presumably high dimensional) parameter space $\pars$.
  Elsewhere, we discuss the computation of this object (\citealt{fengji}),
  and so have many others before us.
  We also have (unpublished) philosophical arguments against calculating
  this $Z$ if you can possibly avoid it, but these are
  outside the scope of this \documentname.
  The point is that \MCMC\ methods will not require that you know $Z$.}
That is, you often know the function $p(\pars\given\data)$ up to a
constant factor; you can compute ratios of the pdf at pairs of points,
but not the precise value at any individual point.

In addition to this normalization-insensitive property of \MCMC, in its
simplest forms it can be run without computing any derivatives or
integrals of the function, and (as we will show below in \sectionname~\ref{sec:MH}) in its simplest
forms it is \emph{extremely easy to implement}.
For all these reasons, \MCMC\ is ideal for sampling posterior pdfs in
the real situations in which scientists find themselves.

Say you are in this situation:
You have a huge blob of data $\data$ (think of this as a vector or
list or heterogeneous-but-ordered collection of observations).
You also have a model sophisticated enough---a probabilistic,
generative model, if you will\note{%
  Briefly, a ``model'' for us is a likelihood function (a pdf for the data given
  model parameters), and a prior pdf over the parameters.
  Because, under this definition, the model can always generate (by
  sampling, say) parameters and parameters can generate (again by
  sampling, say) data, the model is effectively (or actually, if you
  are a true subjective Bayesian) a probability distribution (pdf)
  over all possible data.%
}---that, given a
setting of a huge blob of parameters (again, think of this as a vector or list or heterogeneous-but-ordered collection of values) $\pars$,
  you can compute a pdf for data (or likelihood\note{%
  Techically, $p(\data\given\pars)$
  is only properly a likelihood function when
  we are thinking of the data $\data$ as being fixed, and the parameters $\pars$
  as being permitted
  to vary.}) $p(\data\given\pars)$.
Furthermore, say also that you can write down some kind of informative or
vague prior pdf $p(\pars)$ for the parameter blob $\pars$.
If all these things are true, then---even if you can't compute
anything else---in principle a trivial-to-implement \MCMC\ can give you
a fair sampling of the posterior pdf.
That is, you can run \MCMC\
(for a very long time---see \sectionname~\ref{sec:convergence} for how long)
and you will be left with a set of $K$ parameter-blob settings $\pars_k$ such that the
full set $\setof{\pars_k}_{k=1}^K$ constitutes a fair sampling from
the posterior pdf $p(\pars\given\data)$. We will give some sense of what ``fair'' means
in this context, below.

All that said, and adhering to the traditions of the \project{Data
  Analysis Recipes} project\note{Every entry in the \project{Data
    Analysis Recipes} series begins with a rant in which we argue that
  most uses of the methods in question are not appropriate!}, we are
compelled to note at the outset that \MCMC\ is in fact \emph{over-used}.
Because \MCMC\ provably (under assumptions\note{The assumptions include
  things like:  The algorithm is run ``long enough'', where this phrase
  is undefined (since the convergence requirements are set by precision
  requirements on particular integrals), and that the density that is
  being sampled has some connectedness properties:  There aren't distant islands of finite density
  separated by regions of zero (or exceedingly low) density.}, some of which will be discussed) samples the full
posterior pdf in all of parameter space, many investigators use \MCMC\
because (they believe) it will sample \emph{all} of the parameter space
($\pars$-space).
That is, they are using \MCMC\ because they want to
\emph{search the parameter space for good models}.
This is not a good reason to use \MCMC!
Another bad use case is the following:
Because \MCMC\ samples the parameter representatively, it spends most of
its time near very good models; models that (within the confines of
the prior pdf) do a good job of explaining the data.
For this reason, many investigators are using \MCMC\ because it
effectively \emph{optimizes the posterior pdf}, or, for certain
choices of prior pdf, \emph{optimizes the likelihood}.\note{Of course
  a committed Bayesian would argue that any time you are optimizing a
  posterior pdf or optimizing a likelihood, you \emph{should} be
  sampling a posterior pdf.  That is, for some, the fact that when
  someone ``wants'' to optimize, it is actually useful that they
  choose the ``wrong'' tool, because that ``wrong'' tool gives them
  back something far more useful than the output of any optimizer!}
This is another bad reason!

Both of these reasons for using \MCMC---that it is a parameter-space
search algorithm, and that it is a simple-to-code effective
optimizer---are \emph{not good reasons}.  \MCMC\ is a
\emph{sampler}.  If you are trying to find the optimum of the
likelihood or the posterior pdf, you should use an \emph{optimizer},
not a sampler.  If you want to make sure you search all of parameter
space, you should use a \emph{search algorithm}, not a sampler.  \MCMC\
is good at one thing, and one thing only: Sampling ill-normalized (or
otherwise hard to sample) pdfs.

In what follows, we are going to provide a kind of ``user manual'' or
advice document or folklore capture regarding the use of \MCMC\ for
data analysis.
This will not be a detailed description of multiple \MCMC\ methods
(indeed, we will only explain one method in detail), and it will not
be about the mathematical properties or structure of the method or
methods.
It will be about how to use \MCMC, including diagnosis and
trouble-shooting.

The first couple of \sectionname s will describe what a sampling is, and how
the simplest \MCMC\ method, the Metropolis-Hastings algorithm, can
provide one.
The next few \sectionname s will provide ideas about how to
initialize, tune and operate \MCMC\ methods for good performance.
The last few \sectionname s will provide advice for making decisions
among the myriad \MCMC\ methods implementations, how to implement a good
likelihood function and prior pdf function for inference, and how to
trouble-shoot standard kinds of problems that arise in operating \MCMC\
methods on real problems.
We will leave parenthetical and philosophical matters to the footnotes,
which appear after the main text.

\section{What is a sampling?}\label{sec:sampling}
\nopagebreak
Heuristically, a sampling $\setof{\pars_k}_{k=1}^K$ from some pdf $p(\pars)$
is a set of $K$ values $\pars_k$ that are draws from the pdf.
Heuristically, if an enormous (large $K$) sampling could be displayed
in a finely binned $\pars$-space histogram, the histogram would look---up
to a total normalization---just like the original function $p(\pars)$.

We don't know all the relevant mathematics (measure theory),
but for our purposes here a pdf is
any single-valued (scalar) function that is non-negative everywhere
(the entire domain of $\pars$), and obeys a normalization condition
\begin{eqnarray}
0 &\leq& p(\pars) \quad \mbox{for all $\pars$}
\\ \label{eq:normp}
1 &=& \int p(\pars)\,\dd\pars
\quad ,
\end{eqnarray}
where, implicitly, the integral is over the full domain of $\pars$.
Importantly, although $p(\pars)$ is single-valued, $\pars$ can be
a multi-element vector or list or blob; it can be arbitrarily large
and complicated.
In data-analysis contexts, $\pars$ will often be the full blob of free
parameters in the model.
Implicitly, the integral in \equationname~(\ref{eq:normp}) is high-dimensional; it
has as many dimensions as there are elements or entries or components of $\pars$.
Also, if there are elements or entries or components of $\pars$ that are discrete
(that is, take on only integer values or equivalent),
then along those dimensions the integral becomes a discrete sum.
This latter is a detail to which we return below (briefly, in
  \sectionname~\ref{sec:trouble}).

Given this pdf $p(\pars)$, we can define \emph{expectation values}
$E_{p(\pars)}[\pars]$
for $\pars$ or for any quantity that can be expressed as a function $g(\pars)$
of $\pars$:
\begin{eqnarray}
E_{p(\pars)}[\pars] &\equiv& \int \pars\,p(\pars)\,\dd \pars
\\
E_{p(\pars)}[g(\pars)] &\equiv& \int g(\pars)\,p(\pars)\,\dd \pars
\quad ,
    \label{eq:the-real-integral}
\end{eqnarray}
where again the integrals are implicitly definite integrals over the
entire domain
of $\pars$ (all the parts of $\pars$ space in which $p(\pars)$ is finite) and the
integrals are multi-dimensional if $\pars$ is multi-dimensional.
These expectation values are the mean values of $\pars$ and $g(\pars)$ under
the pdf.  A good sampling---and really this is the definition of a good
sampling---makes the sampling approximation to these integrals
accurate.
With a good sampling $\setof{\pars_k}_{k=1}^K$ the integrals get replaced
with sums over samples $\pars_k$:
\begin{eqnarray}
E_{p(\pars)}[\pars] &\approx& \frac{1}{K}\,\sum_{k=1}^K \pars_k
\\
E_{p(\pars)}[g(\pars)] &\approx& \frac{1}{K}\,\sum_{k=1}^K g(\pars_k)
\quad .
    \label{eq:the-real-samples}
\end{eqnarray}
That is, a sampling is good when any expectation value of interest is
accurately computed via the sampling approximation.
The word ``accurately'' here translates into some kinds of theorems
about limiting behavior; the general idea is that the sampling
approximation becomes exact as $K$ goes to infinity.
The size $K$ of the sampling you need \emph{in practice} will depend
on the expectations you want to compute, and the accuracies you need.

As we noted above (\sectionname~\ref{sec:when}), in the context of \MCMC, we are often
using some \emph{badly normalized} function $f(\pars)$.
This function is just the pdf $p(\pars)$ multiplied by some unknown and
hard-to-compute scalar.
In this case, for our purposes, the conditions on $f(\pars)$ are that it
be non-negative everywhere and have \emph{finite} integral $Z$
\begin{eqnarray}
0 &\leq& f(\pars) \quad \mbox{for all $\pars$}
\\
Z &=& \int f(\pars)\,\dd \pars \label{eq:proper}
\quad .
\end{eqnarray}
And recall that we don't actually know the value of $Z$, but we do know
that it is finite.

When the sampling $\setof{\pars_k}_{k=1}^K$ is of one of these badly
normalized functions $f(\pars)$---as it usually will be---the
sampling-approximation expectation values are the expectation values
under the properly-normalized corresponding pdf, even though you might
\emph{never learn that normalization}.
When we run \MCMC\ sampling on $f(\pars)$, a function that differs from a
pdf $p(\pars)$ by some unknown normalization constant, then the sampling
permits computation of the following kinds of quantities:
\begin{eqnarray}
E_{p(\pars)}[g(\pars)] &\equiv& \frac{\int g(\pars)\,f(\pars)\,\dd \pars}{\int f(\pars)\,\dd \pars}
\\
E_{p(\pars)}[g(\pars)] &\approx& \frac{1}{K}\,\sum_{k=1}^K g(\pars_k)
\label{eq:expectationvalue}
\quad .
\end{eqnarray}
That is, the sampling can be constructed (as we will show below in \sectionname~\ref{sec:MH}) from
evaluations of $f(\pars)$ directly, and it permits you to compute expectation values
without ever requiring you to integrate either the numerator integral
or the denominator integral, both of which are generally
intractable.\note{As we will
  see, the intractability comes from various directions, but one is
  that the dimension of $\pars$ gets large in most realistic situations.
  Another is that the support of $p(\pars)$ tends to be, in normal
  inference situations, \emph{far smaller} than the full domain of
  $\pars$.
  That is, the pdf is at or very close to zero over all but some tiny
  and very hard-to-find part of the space.}

The ``correctness'' of a sampling is defined (above in \sectionname~\ref{sec:sampling}) in terms of its use in
  performing integrals or approximate computation of integrals.
In a deep sense, the \emph{only} thing a sampling is good for is
  computing integrals.
There are many uses of the \MCMC\ sampling, some of them good and some of them bad.
Most of the good or sensible uses will somehow involve integration.

For example, one magical property of a sampling (in a $D$-dimensional
space) is that a \emph{histogram} (a $D$-dimensional histogram) of the
samples (divided, if you like, by the number of samples in each bin
and the bin width\note{We are referring here to the point that if you
  want a histogram of samples to look just like the posterior pdf from
  which the samples are drawn, the histogram (thought of as a step
  function) must integrate to unity.})
looks very much like the pdf from which the samples were drawn.
This is a way to ``reconstruct'' the pdf from the sampling:
Make a histogram from the samples.
Even in this case, the sampling is being used to do \emph{integrals};
  the (possibly odd) idea is that the approximate or effective value of the pdf in each bin
  of the histogram is an average over the bin.
That average is obtained by performing an integral.

Integrals are also involved in finding the mean, median, and any quantiles
  of a pdf.
They are not involved in finding the \emph{mode} of a pdf.
For this reason (and others), in what follows, when we talk about what to report
  about the outcome of your \MCMC\ sampling,
  we will advise in favor of mean, median, and quantiles,
  and we will advise against mode.

Finally---and perhaps most importantly---a magical property of a sampling
  (in a $D$-dimensional space)
  is that if some of your $D$ dimensions are totally uninteresting
  (nuisance parameters, if you will)
  and some of your $D$ dimensions are of great interest,
  the sampling in the full $D$-space is trivially converted into a sampling in the subspace of interest:
You just drop from each $\pars_k$ vector (or blob or list) the dimensions of no interest!
That is, the projection of the sampling to the subspace of interest produces
  a sampling of the marginalized pdf, marginalizing (or projecting) out the
  nuisance parameters.\note{This point about marginalization is, once again,
        a use of the sampling to perform an \emph{integral};
        the marginalized pdf is obtained from the full pdf by an integration.
        The sampling performs this integration automatically.}
That is extremely important for inference,
  where there are always parameters with very different levels of importance to the scientific conclusions.
This point generalizes from a subspace to any function of the parameters;
and it will return again below (\sectionname~\ref{sec:results}).

Although the discussion in this \documentname\ is general, the most
common use of \MCMC\ sampling (for us, anyway) is in probabilistic
inference.
For this reason, we will often refer to the function $f(\pars)$
colloquially as ``the posterior pdf''\note{We will also sometimes
  refer to expectations under $f(\pars)$ as posterior means, and
  medians as median-of-posterior values, and so on.} even though it is
implicitly ill-normalized and might not be a posterior pdf in any
sense.
We will also occasionally assume---just because it is true in
inference---that the function $f(\pars)$ is the product of two
functions, one called ``the prior pdf'' and one called ``the
likelihood''.
Again, this usage is colloquial and is only strictly correct in
inference contexts with proper inputs.
That the function $f(\pars)$ can be thought of as a product of a
prior pdf and a likelihood is only necessary for what follows in
the context of advanced sampling techniques like tempering or
nested sampling, both mentioned briefly below (\sectionname~\ref{sec:methods}).

\begin{problem}\label{prob:simple-stats}
Look up (or choose) definitions for the mean, variance, skewness,
and kurtosis of a distribution.
Also look up or compute the analytic values of these four statistics
for a top-hat (uniform) distribution.
Write a computer program that uses some standard package (such as
\project{numpy}\note{There isn't a full citation for this package but there is \cite{numpy}.})
to generate $K$ random numbers
$x$ from a uniform distribution in the interval $0<x<1$.
Now use those $K$ numbers to compute a sampling estimate of the
mean, variance, skewness, and kurtosis (four estimates; look up
definitions as needed).
Make four plot of these four estimates as a function of $1/K$ or
perhaps $\log_2 K$, for $K=4^n$ for $n=1$ up to $n=10$ (that is,
$K=4$, $K=16$, and so on up to $K=1048576$).
Over-plot the analytic answers.
What can you conclude?
\end{problem}

\section{Metropolis--Hastings MCMC}\label{sec:MH}
\nopagebreak
The simplest algorithm for \MCMC\ is the Metropolis--Hastings algorithm (\MH\ \MCMC).\note{%
  There are many claims that this algorithm is incorrectly named,
  with claims that it should be credited to Enrico Fermi or Stan Ulam.
  We don't have any opinions of this matter, but encourage the reader to follow this up.
  The original paper is \citet{metropolis}, and there are a few sketchy historical notes
  in \citet{geyer}.}
It is so simple that we recommend that any reader of this document
  who has not previously implemented the algorithm take a break at the end
  of this \sectionname\ and implement it forthwith, in a short piece of computer code,
  in the context of some simple problems.\note{%
    We request this in a trivial case below in
    \problemname~\ref{prob:MH} and the following problems in this
    \sectionname.
    We make the same request in a less trivial context in our
    model-fitting screed (\citealt{fittingaline},
    \problemname~6 and 7), where we ask the Reader to implement \MCMC\ for
    a useful mixture model.}

The \MH\ \MCMC\ algorithm requires two inputs.
The first is a handle to the function $f(\pars)$ that is the function to be sampled,
  such that the algorithm can evaluate $f(\pars)$ for any value of the parameters $\pars$.
In data-analysis contexts,
  this function would be the prior $p(\pars)$ times the likelihood $p(\data\given\pars)$
  evaluated at the observed data $\data$.
The second input is a handle to a proposal pdf function $q(\pars'\given\pars)$ that can deliver samples,
  such that the algorithm can draw a new position $\pars'$ in the parameter space
  given an ``old'' position $\pars$.
This second function must meet a symmetry requirement (detailed
  balance) we discuss further below.
It permits us to random-walk around the parameter space in a fair way.

The algorithm\note{Technically this is the \emph{Metropolis} algorithm
  rather than the Metropolis--Hastings algorithm.  The difference is
  that in the Metropolis--Hastings algorithm, we permit the proposal
  distribution to disobey detailed balance but correct the
  accept--reject step to recover detailed balance.} is the
following: We have generated some set of samples, the most recent of
which is $\pars_k$ To generate the next sample $\pars_{k+1}$ do the
following:
\begin{itemize}
\item Draw a proposal $\pars'$ from the proposal pdf $q(\pars'\given\pars_k)$.
\item Draw a random number $0<r<1$ from the uniform distribution.
\item If $f(\pars') / f(\pars_k) > r$ then $\pars_{k+1} \leftarrow \pars'$;
      otherwise $\pars_{k+1} \leftarrow \pars_k$.
\end{itemize}
That is, at each step, either a new proposed position in the parameter
  space gets accepted into the list of samples or else the previous sample
  in the parameter space \emph{gets repeated}.
The algorithm can be iterated a large number $K$ of times to produce $K$ samples.

Why does this algorithm work?  The answer is not absolutely
trivial\note{There is a very mathematical discussion in \citet{geyer}
  that we do not fully understand.  There is a more heuristic answer
  on \project{Wikipedia} that we are happier with!},
but there are two components to the argument:
The first is that the Markov process delivers a unique stationary
distribution.
The second is that that stationary distribution is proportional to the
density function $f(\pars)$.

This algorithm---and indeed any \MCMC\ algorithm---produces a
  \emph{biased random walk} through parameter space.
It is a random walk for the same reason that it is ``Markov'':
The step it makes to position $\pars_{k+1}$ depends only on the state
  of the sampler at position $\pars_k$ (and no previous state).
It is a biased random walk, biased by the acceptance algorithm
  involving the ratios of function values; this acceptance rule biases
  the random walk such that the amount of time spent in the neighborhood
  of location $\pars$ is proportional to $f(\pars)$.
Because of this local (Markov) property, the nearby samples are not
  independent; the algorithm only produces fair samples in the limit of
  arbitrary run time, and two samples are only independent when they are
  sufficiently separated in the chain (more on this below in \sectionname~\ref{sec:convergence}).

The principal user-settable knob in \MH\ \MCMC\ is the proposal
  pdf $q(\pars'\given\pars)$.
A typical choice is a multi-variate Gaussian distribution for $\pars'$
  centered on $\pars$ with some simple (diagonal, perhaps) variance
  tensor.
We will discuss the choice and tuning of this proposal distribution
  below (\sectionname~\ref{sec:tuning} and \sectionname~\ref{sec:trouble}).

Importantly---for the algorithm given above to work correctly---the
  proposal pdf must satisfy a ``detailed-balance'' condition\footnote{This
  detailed-balance condition in \equationname~(\ref{eq:detailed-balance})
  requires implicitly that the function $q(\pars'\given\pars)$ is also
  properly normalized. That is, that it integrates (over $\pars'$) to unity
  for any setting of $\pars$. That is an extremely technical point, but stands
  as a reminder that you don't want to mess with detailed balance casually!};
  it must have the property that
\begin{eqnarray}
q(\pars'\given\pars) &=& q(\pars\given\pars')
\label{eq:detailed-balance}\quad ;
\end{eqnarray}
that is, it must be just as easy to go one way in the parameter space
  as the other.
You can break this property, if you like, and then adjust the
  acceptance condition accordingly (which is truly the Metropolis--Hastings algorithm), but we \emph{do not recommend this}
  except under the supervision of a trained professional.\note{Actually,
  our own favorite method, \project{emcee}, has a proposal pdf that does
  violate detailed balance in this way and has a compensated
  acceptance probability.  Read the paper \cite{emcee} for more details.}
The reason is:  It is one thing to draw samples from $q(x'\given x)$.
It is another thing to correctly write down $q(x'\given x)$.
  If you violate detailed balance in $q(x'\given x)$ then you have to be
  able to \emph{both} draw from \emph{and} write down $q(x'\given x)$
  and you are subject to a set of new possible bugs for your code.
If the detailed balance condition frightens you (as it frightens us),
  then just stick with pdfs that are symmetric in $\pars'$ and $\pars$,
  like the Gaussian, or a centered uniform distribution, and forget about
  it.

In what follows, when we discuss tuning of the proposal pdf, we will
  often cycle through the parameter blob components and propose changes
  in only one dimension or element or component at a time.
That is, at step $k+1$ you might only propose a one-dimensional move
  in the $i$th dimension, not in all $D$ dimensions of the parameter
  space.
That won't change anything significant; but if you are implementing
  this right now you might want to do that.
It will help us tune the proposal pdf (\sectionname~\ref{sec:tuning})
and diagnose problems (\sectionname~\ref{sec:trouble}).

One note to make here
  is that you must \emph{propose}
  in the same coordinates (parameterization or transformation of parameters)
  as that in which your priors are \emph{specified},
  or else multiply in a (possibly nasty) Jacobian.
That is, if your prior is ``flat in $\pars$''
  but you find it easier to run your sampler by proposing steps in $\ln \pars$,
  if you don't modify your acceptance probability by the Jacobian,
  your \emph{real} prior won't be flat in $\pars$.
These think-os can get subtle;
  the best way to avoid them
  is to write your code and your likelihood function
  and your prior pdf function and your proposal distribution
  all in precisely the same parameterization.
We will return to this point below in our comments on testing (\sectionname~\ref{sec:trouble}).
The point is that if you \emph{do} change coordinates between
  the statement of your priors and the specific variables you
  step in, you have to introduce Jacobian factors.

As we discuss briefly below (\sectionname~\ref{sec:methods}),
  there are many \MCMC\ methods more advanced than M--H.
However, they all share characteristics with M--H
  (initialization, tuning, judging convergence),
  such that it is very valuable to understand \MH\ well before using anything more advanced.
Furthermore, a scientist new to \MCMC\ benefits enormously from
  building, tuning, and using her or his own \MCMC\ software.
One piece of advice we give then,
  to the new user of \MCMC,
  is to code up, tune, and use a \MH\ \MCMC\ sampler for a scientific project.
A huge amount is learned in doing this,
  and it is very often the case that the home-built \MH\ \MCMC\ does everything needed;
  you often don't need any more advanced tool.
Only after you have concluded that your home-built \MH\ \MCMC\ sampler
  is not suited to your project
  (or not developed properly into a properly versatile software package)
  should you download and start to use any professionally developed alternative.
That is, even if---in the end---%
  you want to leave the sampling code to the experts and run an industrial-strength code,
  it is still valuable to build your own,
  given the simplicity of the algorithm,
  and given the intuition you gain by doing it (at least once) yourself.\note{%
    We also make this point in a previous piece (\citealt{fittingaline});
    the ambitious reader will find in that piece a solid example project
    for using \MCMC\ in a real context.}

Because many problems involve a huge amount of dynamic range in the
density function $f(\pars)$, and we like to avoid underflows and overflows
(in, say, ratios computed for accept--reject steps), it is often advisable
to work in the (natural) logarithm of the density rather than the density.
When working in logarithmic density,
the accept--reject step would change from a comparison of a random
deviate with a ratio of probabilities to comparison of the log of a
random deviate with a difference of log probabilities.
That is, the accept--reject step becomes:
\begin{itemize}
\item If $\ln f(\pars') - \ln f(\pars_k) > \ln r$ then $\pars_{k+1} \leftarrow \pars'$;
      otherwise $\pars_{k+1} \leftarrow \pars_k$.
\end{itemize}
This protects you from underflow, but exposes you to some (negative)
infinities, if you end up taking the logarithm of a zero.  It is
important to write your code to be infinity-safe.\note{Most code will
  be infinity safe if written in the form in this paragraph. However,
  there can be issues if \emph{both} $f(\pars')$ \emph{and}
  $f(\pars_k)$ are negative infinity; in (present-day) Python this
  will return a \texttt{NaN} rather than an infinity or zero. That's a
  problem and is a case that should get caught (probably way before
  the accept--reject step!).}

\begin{figure}[!htbp]
\begin{center}
\includegraphics[width=0.5\textwidth]{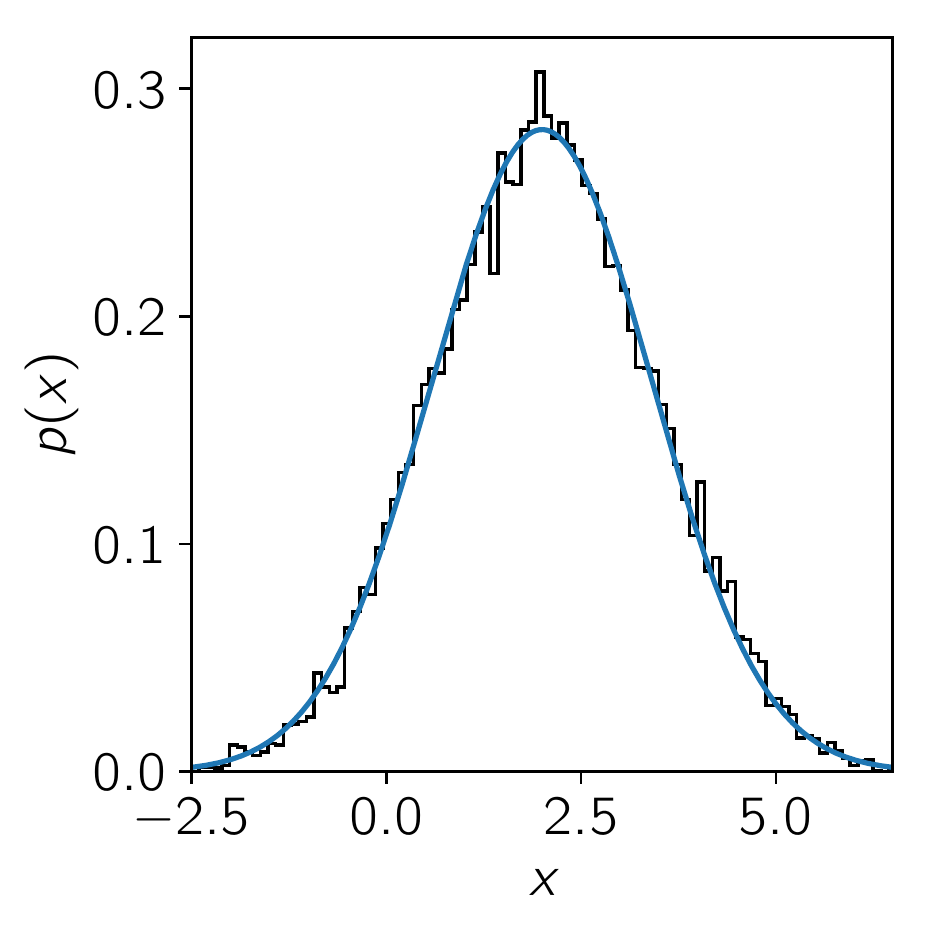}
\end{center}
\caption{Solution to \problemname~\ref{prob:MH}:
\MCMC\ samples from a Gaussian (black) and the true distribution (blue).}
\label{fig:MH}
\end{figure}

\begin{problem}\label{prob:MH}
In your scientific programming language of choice, write a very simple
M-H \MCMC\ sampler.
Sample in a single parameter $x$ and give the sampler as its density
function $p(x)$ a Gaussian density with mean 2 and variance 2.
(Note that variance is the \emph{square} of the standard deviation.)
Give the sampler a proposal distribution $q(x'\given x)$ a Gaussian
pdf for $x'$ with mean $x$ and variance 1.
Initialize the sampler with $x=0$ and run the sampler for more than $10^4$ steps.
Plot the results as a histogram, with the true density over-plotted sensibly.
The resulting  plot should look something like \figurename~\ref{fig:MH}.
\end{problem}

\begin{figure}[!htbp]
\begin{center}
\includegraphics[width=0.5\textwidth]{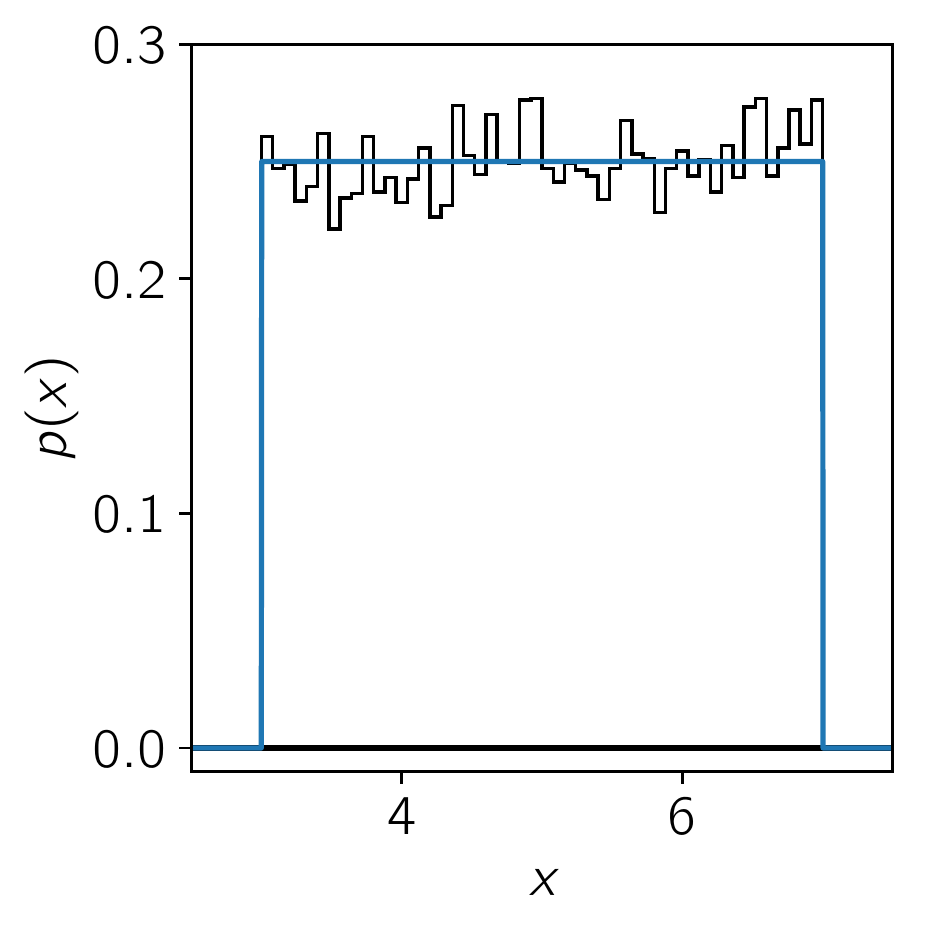}
\end{center}
\caption{Solution to \problemname~\ref{prob:MH2}:
\MCMC\ samples from a uniform distribution (black) and the true distribution
(blue).}
\label{fig:MH2}
\end{figure}

\begin{problem}\label{prob:MH2}
Re-do \problemname~\ref{prob:MH} but now with an input density
that is uniform on $3<x<7$ and zero everywhere else.
The plot should look like \figurename~\ref{fig:MH2}.
What change did you have to make to the initialization, and why?
\end{problem}

\begin{figure}[!htbp]
\begin{center}
\includegraphics[width=0.7\textwidth]{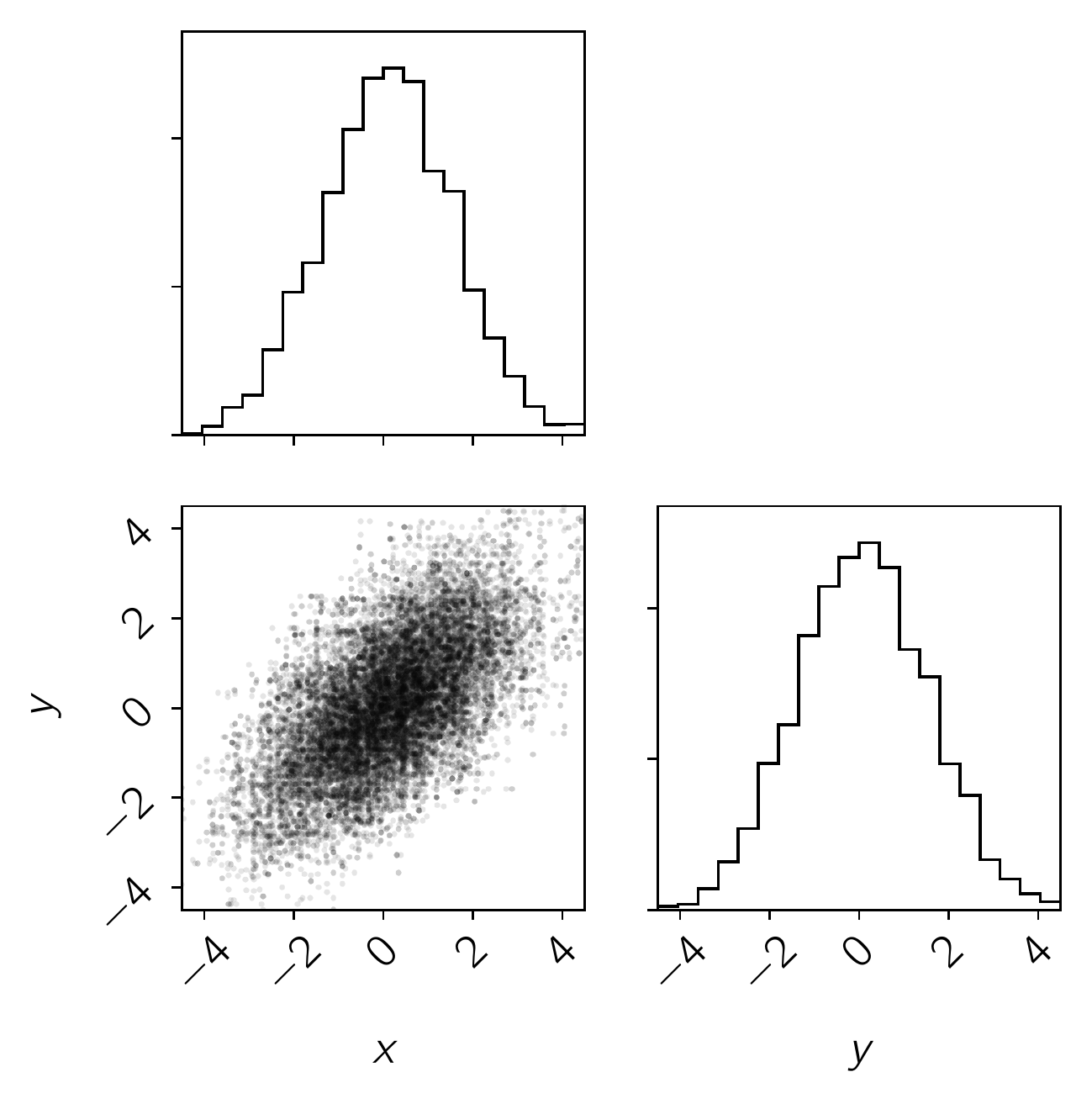}
\end{center}
\caption{Solution to \problemname~\ref{prob:twod}a:
\MCMC\ samples from a two-dimensional Gaussian (scatter plot) and the
one-dimensional marginalized distributions (histograms).}
\label{fig:twod-a}
\end{figure}

\begin{problem}\label{prob:twod}
Re-do \problemname~\ref{prob:MH} but now with an input density
that is a function of two variables $(x, y)$.
For the density function use two different functions.
\emph{(a)} The first density function is a covariant two-dimensional Gaussian
density with variance tensor
\begin{eqnarray}
V &=& \left[\begin{array}{cc} 2.0 & 1.2 \\ 1.2 & 2.0 \end{array}\right]
\quad.
\end{eqnarray}
\emph{(b)} The second density function is a rectangular top-hat function that
is uniform on the joint constraint $3<x<7$ and $1<y<9$ and zero everywhere else.
For the proposal distribution $q(x', y'\given x, y)$ a two-dimensional Gaussian
density with mean at $[x, y]$ and variance tensor set to the two-dimensional
identity matrix.
Plot the two one-dimensional histograms and also a two-dimensional scatter
plot for each sampling.
\figurename~\ref{fig:twod-a} shows the expected results for the Gaussian.
Make a similar plot for the top-hat.
\end{problem}

\begin{problem}\label{prob:MH-sigma}
Re-do \problemname~\ref{prob:twod}a but with different values for
the variance of the proposal distribution $q(x'\given x)$.
What happens when you go to very extreme values (like for instance
$10^{-1}$ or $10^2$)?
\end{problem}

\begin{problem}\label{prob:prop-mean}
Why, in all the previous problems, did we give the proposal
distributions $q(x'\given x)$ a mean of $x$?
What would be bad if we hadn't done that?
Re-do \problemname~\ref{prob:twod}a with a proposal $q(x'\given x)$ with a
stupidly shifted mean of $x + 2$ and see what happens.
\emph{Bonus points:} Modify the acceptance--rejection criterion to deal with
the messed-up $q(x'\given x)$ and show that everything works once
again.
\end{problem}

\section{Likelihoods and priors}\label{sec:likelihood}
\nopagebreak
\MCMC\ is used to obtain samples $\pars_k$ from a pdf $p(\pars)$,
  given a badly normalized all-positive function $f(\pars)$
  that is different from the pdf by an unknown factor $Z$.
In the context of data analysis,
  \MCMC\ is usually being used to obtain samples $\pars_k$
  that are \emph{parameter values} $\pars$ for a probabilistic model of some data.
That is, \MCMC\ is sampling a pdf \emph{for the parameters}.

Ideally, if you are using \MCMC\ for inference, your code should input
to the \MCMC\ function or routine a probability function which is a
product of a prior times a likelihood.
As we noted above---in terms of implementation---it is usually advisable to work in the
logarithm of the density function, so the function input to the \MCMC\ code
would be called something like \code{ln_f()}.
This function \code{ln_f()} internally would compute and return the
sum of a log prior pdf \code{ln_prior()} and a log likelihood function
\code{ln_likelihood()}.

If you are using ``flat'' (improper) priors, the \code{ln_prior()}
function can just return a zero no matter what the parameters.
If it is flat with bounds (that is, proper), the \code{ln_prior()}
should check the bounds and return \code{-Inf} values when the parameter
vector is out of bounds.
The pseudo-code for the \code{ln_f()} function should look something
like this:
\begin{verbatim}
def ln_f(pars, data):
    x = ln_prior(pars)
    if not is_finite(x):
        return -Inf
    return x + ln_likelihood(data, pars)
\end{verbatim}
This pseudo-code ensures that you only compute the likelihood when the
parameters are within the prior bounds; it assumes implicitly that the
prior pdf is easier to compute than the likelihood.
If you find yourself in the opposite regime, adjust accordingly.
Of course the above pseudo-code presumes that when you perform the
accept--reject step of the \MCMC\ method, the programming language handles
properly the \code{-Inf} values.\note{If you are working in a language
that doesn't have an \code{Inf} or doesn't evaluate comparisons correctly
when the \code{Inf} appears, you might have to write some case code, and
have your \code{ln_f} function return a value \emph{and} some kind of flag
which indicates ``zero probability'' or $f(\pars)=0$.}

\MCMC\ \emph{cannot sample a likelihood} (which is a probability \emph{for the data} given
parameters).\note{Well, technically, \MCMC\ \emph{can} be used to sample a likelihood function,
  but the samples would be samples of possible \emph{data} not possible \emph{parameters}.
  The purist might say that even this is not sampling a likelihood function,
  because you should only call it a ``likelihood function'' in contexts in which you are
  treating the data as fixed (at the true data values) and the parameters as variable.
  In this context, the likelihood function is \emph{not} a pdf for anything,
  so it can't be sampled.}
Despite this, in many cases,
  data analysts believe they are sampling the likelihood.
This is because (we presume) they have put a likelihood function
  (or log-likelihood function)
  in as the input to the \MCMC\ code, where the probability function
  (or log-probability function) should go.
Then is it the likelihood that is being sampled?
No, not really;
  it is a posterior probability that is directly proportional to the likelihood function.
That is, it is a posterior probability for some implicit (and improper) ``flat'' priors.

It should be outside of the scope of this document to note here that
it is a good idea to have proper priors.\note{This point---that you
  should be using proper priors---is just basic Bayesian good
  practice, often violated but never for good reasons.}
Proper priors obey the integral constraint (\ref{eq:proper}),
with $Z$ finite.
It isn't a requirement that priors be proper for the posterior to
be proper, so many investigators and projects violate this rule.
This is outside the current scope, except for the following point:
It is a great functional test of your sampler and your data analysis
setup to take the likelihood function to a very small power (much less
than one) or multiply the log-likelihood by a very small number (much
less than one) and check that the sampler samples, properly and
correctly, the prior pdf.
This test is only possible if the prior pdf is proper.

\begin{problem}\label{prob:improper}
Run your M-H \MCMC\ sampler from \problemname~\ref{prob:MH}, but now
with a density function that is precisely unity \emph{everywhere}
(that is, at any input value of $x$ it returns unity).  That is, an
improper function (as discussed in \sectionname~\ref{sec:likelihood}).
Run it for longer and longer and plot the chain value $x$ as a
function of timestep.  What happens?
\end{problem}

\begin{problem}\label{prob:fittingaline}
For a real-world inference problem, read enough of
\citet{fittingaline} to understand and execute Exercise~6 in that
document.
\end{problem}

\begin{problem}\label{prob:logprior}
Modify the sampler you wrote in \problemname~\ref{prob:MH} to take
steps not in $x$ but in $\ln x$.  That is, replace the Gaussian
proposal distribution $q(x'\given x)$ with a Gaussian distribution in
$\ln x$ $q(\ln x'\given \ln x)$, but make no other changes.
By doing this, you are no longer sampling the Gaussian $p(x)$ that you were in
\problemname~\ref{prob:MH}.
What about your answers change?  What distribution are you sampling now?
Compute the analytic function that you have sampled from~--~this will no longer
be the same $p(x)$~--~and over-plot it on your histogram.
\end{problem}

\begin{figure}[!htbp]
\begin{center}
\includegraphics[width=0.5\textwidth]{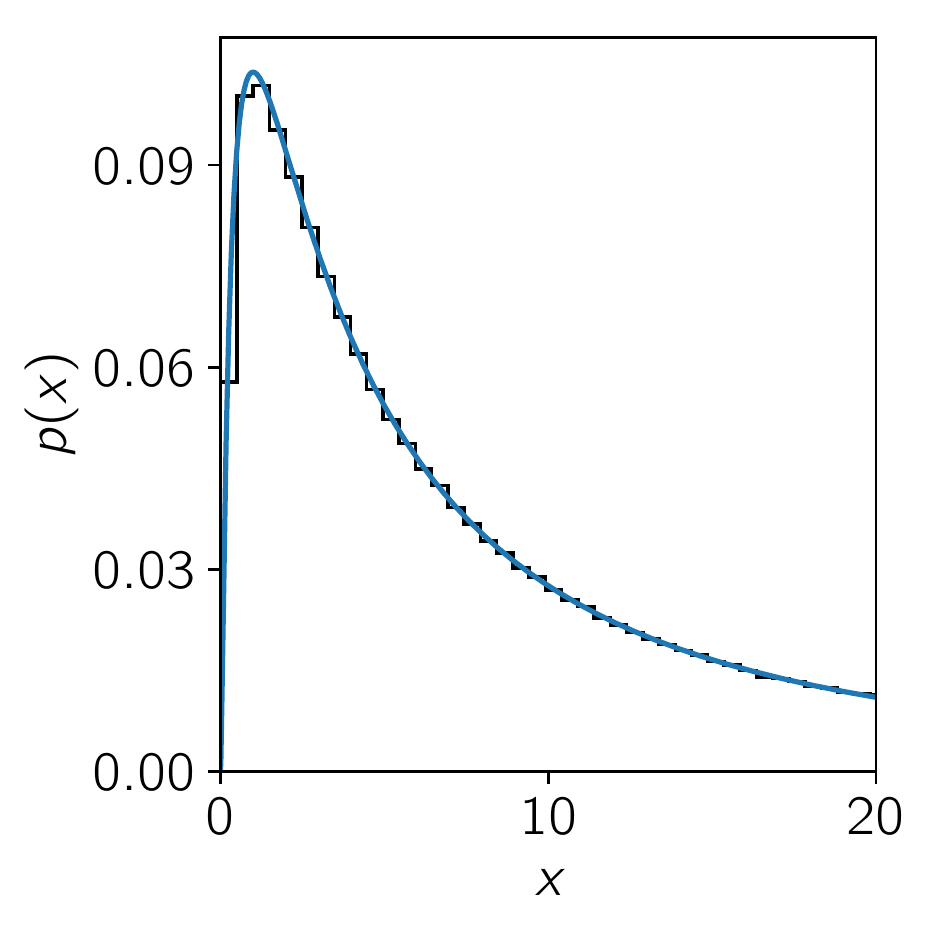}
\end{center}
\caption{Solution to \problemname~\ref{prob:logprior}:
The black histogram shows the \MCMC\ samples and the blue curve is the analytic
target density.}
\label{fig:logprior}
\end{figure}

\Needspace{7\baselineskip}
\section{Autocorrelation \& convergence}\label{sec:convergence}%
\nopagebreak%
A key question for an \MCMC\ operator---\emph{the} key question in some
sense---is \emph{how long to run} to be sure of having reliable
results.
It is disappointing and annoying to many that there is no extremely
simple and reliable answer to this question.\note{There is very good
  coverage of most of the points in this \sectionname, and more, in a
  set of lecture notes by Patrick Lam up at
  \url{http://www.people.fas.harvard.edu/~plam/teaching/methods/convergence/convergence_print.pdf}.}

The reason that there is no simple answer is that you can't really
ever know that you have sampled the full posterior pdf, and the reason
for \emph{that} is that if you \emph{could} know that, you would also
be able to solve the famously difficult discrete optimization
problem.\note{Famously, there is ``no free lunch''
  \citep{nofreelunch}: You can't find the global optimum of a general
  optimization problem without exhaustive search of all possibilities.
  This is very closely related---somehow---to the difficulty of sampling.\label{note:nofreelunch}}
Qualitatively, you can consider the scenario where you have two modes---two
separated regions of substantial total probability---in the posterior pdf that
are both important but separated by a large region of low probability.
If you are using a simple sampler, you could sample one of these modes
very well but the sampler will take effectively infinite time to find
the other mode.
In most real problems, the situation is much worse, with an unknown number of
modes, and knowing that you have a complete and representative sampling
is effectively impossible.

In this context,
  it is \emph{simply not fair} to require an \MCMC\ run to have fully and completely
  sampled the posterior pdf,
  at least not in any provable sense.
The question of whether a sampling is definitely converged to a representative sampling
  of the posterior pdf is actually \emph{outside the domain of science},
  because the domain of science is the domain of questions answerable by scientists.
That is, unless you have some bounds on the support and smoothness of
  the posterior pdf,
  you can \emph{never know} that you have correctly sampled the posterior pdf,
  despite many statements in the literature to the contrary.\note{%
For example, you can (almost) never say that your chains are
definitely converged, or that you have the posterior pdf correct to
some given level of accuracy.  The reason is related to the above-mentioned ``no free
lunch'' theorem of discrete optimization (see \notename~\ref{note:nofreelunch}): In most problems there is no
way you could have searched your parameter space finely enough to know
this.  There are some exceptions of course.  In one kind of exception,
your problem is convex or Gaussian, and you can analytically show that
there is exactly one mode in the density and where it is.  Of course
in these cases you rarely need \MCMC\ to solve your problems!  In
another kind of exception, you can say something about the finite
width or shape of the modes of the likelihood function (and prior
pdf).  For example, sometimes the Cram\'er--Rao bound tells you that
modes of the density must be smoother than some finite amount.  Then
an exhaustive search of parameter space followed by \MCMC\ can in
principle have provable properties.  But, as we say, these situations
are rare.}
The upshot of this is that we don't and can't require any kind of absolute convergence;
  indeed, it would be impossible even to test for it.

In this pragmatic situation---the Real World, as it is called---we have to rely on heuristics.
Heuristically, you have sampled long enough when you can see
that the (or each) walker has traversed the high-probability parts of
the parameter space many times in the length of the chain.
Or, equivalently, you have sampled long enough when the first half of
the chain shows very much the same posterior pdf morphology as the
second half of the chain, or indeed as any substantial subset.
Or, relatedly, you have sampled long enough when different walkers initialized
differently and run with different random number seeds create posterior inferences
that are substantially the same.

The above heuristics can be made more precise in terms of the amount
of deviation one expects between the means and variances, say, of two
disjoint subsets of the chain.
The premier tool for making this heuristic precise, however, is to
look at the \emph{``integrated autocorrelation time''} of the chain.
In general, when one has a sequence $(\pars_1, \pars_2, \theta_3, \cdots)$
generated by a Markov Process in the $\pars$ space, nearby points in
the sequence will be similar, but sufficiently distant points will not
``know about'' one another;
the autocorrelation function measures this.

More formally, as discussed in \sectionname~\ref{sec:sampling}, the goal of
\MCMC\ sampling is to compute integrals of the form given in
\equationname~\ref{eq:the-real-integral} using the Monte Carlo approximation
in \equationname~\ref{eq:the-real-samples}.
The Monte Carlo error introduced by this approximation is proportional to
$\sqrt{\tau_\mathrm{int}/N}$ where $\tau_\mathrm{int}$ is the
integrated autocorrelation time and $N$ is the total number of
samples of $p(\pars)$.
In other words, $\tau_\mathrm{int}$ is the number of steps required for the
chain to produce an \emph{independent} sample.

A sampler with a smaller integrated autocorrelation time is better; you have
to do fewer $f(\pars)$ calls per independent sample, and you have to
run less time to get accurate sampling-based integral estimates.
A sampler that takes an independent sample every time would have an
autocorrelation time of unity, which is the best possible value; this
optimal sampling is only possible for problems where the sampling is
analytic (for example if the posterior is perfectly Gaussian with
known mean and variance).
In general, the best sampler will be different for different problems, and we
will (below; \sectionname~\ref{sec:methods}) tune the samplers we have to do
the best on the problems we have; this tuning will also be problem-specific.
There are many heuristic bases on which different samplers might be
compared or tuned, but fundamentally it is lower autocorrelation time
that separates good samplers from bad ones and is the ultimate basis
on which we compare performance.\note{%
  We sometimes see \MCMC\ methods compared according to burn-in times,
  which, for one, depends strongly on initialization, and, for two,
  depends strongly on tuning and dynamics.  Similarly we often see
  comparisons in terms of acceptance ratio.  In principle the only
  question is how precise are one's inferences given a certain amount
  of computation.  This is set by the autocorrelation time and (pretty
  much) nothing else, for reasonably converged chains.  In the real
  world, of course, one must add to the CPU time the investigator time
  spent tuning the method (and thinking about initialization) but
  there is never (to our knowledge) a condition in which burn-in time
  is the dominant consideration in choosing an \MCMC\ method.}

We won't go into details about how to estimate $\tau_\mathrm{int}$, but it is
notoriously difficult: It is a two-point statistic, and two-point statistics
are much harder to estimate than one-point statistics.\note{See: All of
cosmology!}
If you are not a hard-core user of \MCMC, and if all you want is
heuristic indicators of convergence, then what you should take from this
section is that the autocorrelation time is involved in variance estimates,
but that it is hard to estimate.
If you want to think about autocorrelation estimation, more detailed
references can be found elsewhere.\note{%
    A canonical reference is a set of lecture notes by Alan Sokal
    \citep{sokal}.
    Another good reference is a blog post by one of us (DFM) that can be found
    at \url{http://dfm.io/posts/autocorr}.}

Besides estimating the integrated autocorrelation time, another simple and
sensible test of convergence is the Gelman--Rubin
diagnostic\note{\citet{gelmanrubin}.}, which compares the variance (in one
parameter, or your most important parameter, or all parameters) \emph{within}
a chain to the variance \emph{across} chains.
This requires running multiple chains, and looking at the empirical
variance of the parameter away from its mean within each individual
chain, and comparing it to the variance in the mean of that parameter
across chains, inflated to be a mean per-sample variance.
What Gelman and Rubin do with these variances specifically is
sensible, but the important question of convergence is whether, as the
chains get longer, these two variances asymptotically reach stable
values, and that those two values agree.
The Gelman--Rubin diagnostic is related to the autocorrelation time,
in that it will only deliver success when the chains are much longer
than an autocorrelation time.

Finally one last point about convergence:
Since all convergence tests are fundamentally heuristic, it is useful
to just make the heuristic visualization of the samples $\pars_k$ as a
function of $k$, in the order they were generated.
The chain is likely to be converged only if the random-walk process
crossed the domain of $\pars$ fully many times during the \MCMC\ run.
Often some parameter directions are much worse than others; it is
worth looking at the chain in all parameter directions.

\begin{problem}\label{prob:convergence}
Re-do \problemname~\ref{prob:MH} but now look at convergence:
Plot the $x$ chain as a function of timestep.  Also split the chain
into four contiguous segments (the first, second, third, and fourth
quarters of the chain).  In each of these four, compute the empirical
mean and empirical variance of $x$.  What do you conclude about
convergence from these heuristics?
\end{problem}

\begin{figure}[!htbp]
\begin{center}
\includegraphics[width=0.5\textwidth]{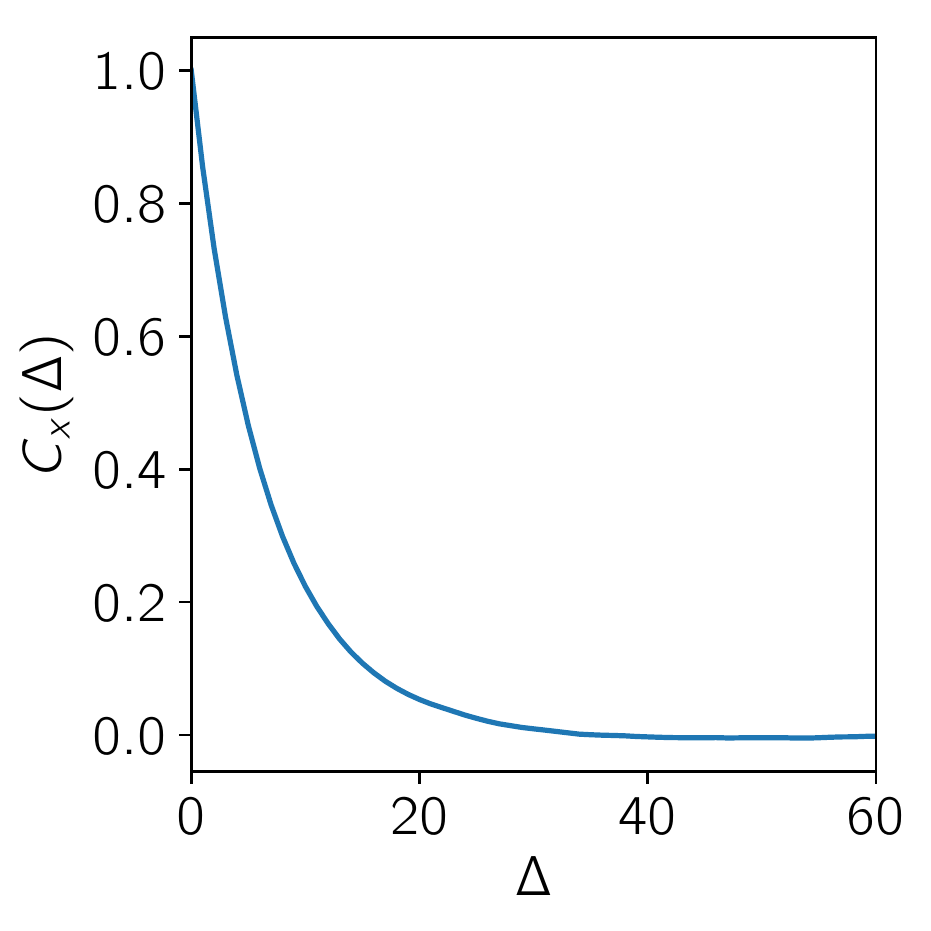}
\end{center}
\caption{Solution to \problemname~\ref{prob:estimatetau}:
The autocorrelation function of the chain from \problemname~\ref{prob:MH}.}
\label{fig:estimatetau}
\end{figure}

\begin{problem}\label{prob:estimatetau}
Write a piece of code that computes the empirical autocorrelation
function.
You will probably want to speed this computation up by using a fast Fourier
transform\note{%
The calculation of the autocorrelation function can be seen as a convolution
and it can, therefore, be computed using the fast Fourier transform in
$\mathcal{O}(N\,\log N)$ operations instead of $\mathcal{O}(N^2)$ for a naive
implementation.}.
Run this on the chain you obtained from \problemname~\ref{prob:MH}.
Plot the autocorrelation function you find at short lags ($\Delta < 100$).
This plot should resemble \figurename~\ref{fig:estimatetau}.
\end{problem}

\begin{figure}[!htbp]
\begin{center}
\includegraphics[width=0.5\textwidth]{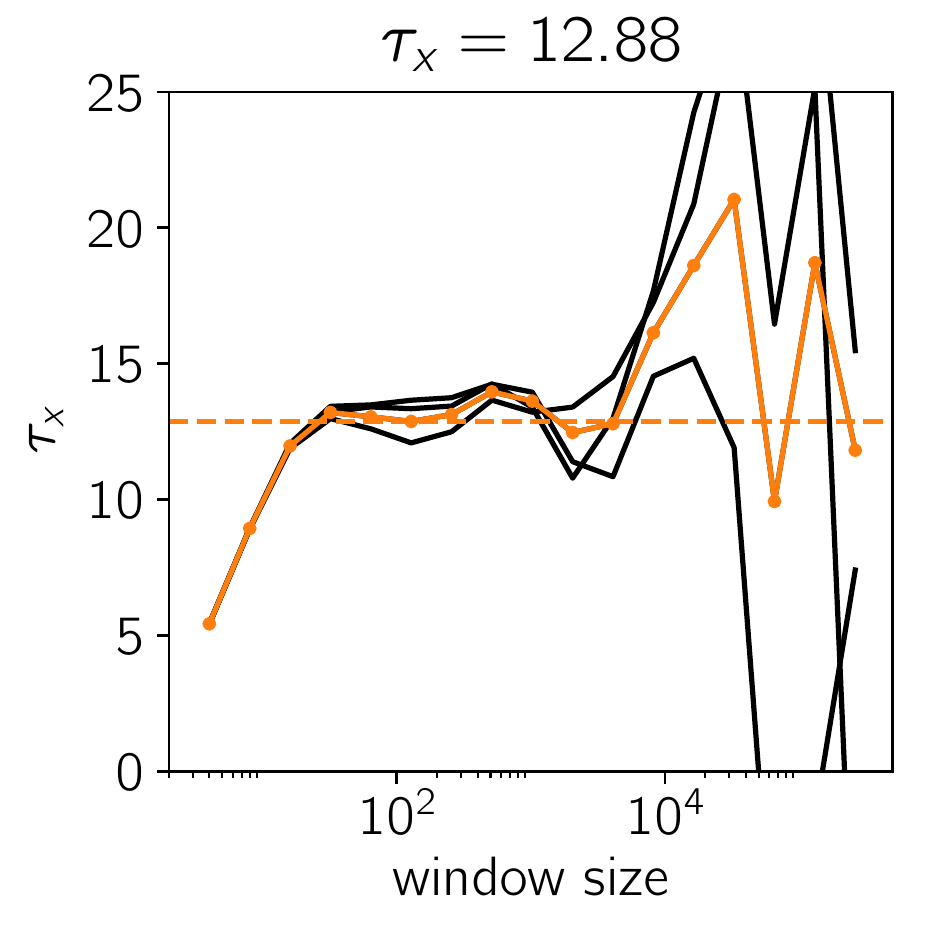}
\end{center}
\caption{Solution to \problemname~\ref{prob:itertau}:
Estimates of the integrated autocorrelation time of different segments of the
\MCMC\ chain from \problemname~\ref{prob:MH} (black lines) and for the full
chain (orange line) as a function of window size $M$.
The ``optimized'' value computed using an iterative procedure is overplotted
as a dashed line and its value is listed in the title.}
\label{fig:itertau}
\end{figure}

\begin{problem}\label{prob:itertau}
Write a piece of code that estimates the integrated autocorrelation time for a
chain of samples using an estimate of the autocorrelation function and a given
``window'' size $M$ (see \citealt{sokal}).
Plot the estimated $\tau$ as a function of $M$ for several contiguous segments
of the chain and overplot the sample function based on the full chain.
What can you conclude from this plot?
Implement an iterative procedure for automatically choosing $M$.\note{%
The recipe given on page 16 of Sokal's notes \citep{sokal} might be helpful.
Note that the definition of $\tau$ that we adopt is twice the value used by
Sokal.}
Overplot this estimate on the plot of $\tau(M)$ and the result should look
like \figurename~\ref{fig:itertau}.
\end{problem}

\section{Tuning}\label{sec:tuning}
\nopagebreak
Most \MCMC\ methods make use of something like a ``proposal distribution'',
  which determines what kinds of steps the walker can take
  as it random-walks through the parameter space.
This is the function we called $q(\pars'\given\pars)$ in \sectionname~\ref{sec:MH}.
The user generally has a lot of control over what this proposal distribution might be,
  and how to choose the parameters.
For example, in a $D$-dimensional parameter space, 
  a zero-mean but anisotropic Gaussian (normal distribution)
  is often used to draw offsets for the walker to take from point to point.
Even within this choice (which is of one function among many possibilities),
  there are $D\,(D+1)/2$ parameters in the $D\times D$ symmetric, positive definite covariance matrix
  to set ``by hand''.
How to choose this function and set these parameters?

The key idea here is that if the proposal distribution is too narrow---%
  it proposes steps too small---%
  almost all steps will be accepted (recall the acceptance-rejection step from \sectionname~\ref{sec:MH})
  but it will take a long time to move anywhere because of timidity.
If the proposal distribution is too wide---%
  it proposes steps too large---%
  the moves will cover parameter space easily,
  but almost no steps will be accepted;
  it will tend to jump to much lower probability regions.
There is a Goldilocks step size (proposal distribution root-variance) that is ``just right''.
In one dimension this might be easy to find,
  but, as we say, in large numbers of dimensions,
  there is a lot of freedom in choosing the parameters of the distribution.

In terms of long-term computational efficiency,
  the only scalar that makes sense to optimize,
  when choosing proposal distribution (or, loosely speaking, step size),
  is the autocorrelation time, described above (\sectionname~\ref{sec:convergence}).
The optimal step size is the step size that makes for the shortest autocorrelation time.
This is easy to state, but hard to use in practice;
  being a second-order statistic of the \MCMC\ chain,
  the autocorrelation time is a hard thing to measure without a lot of data,
  so it is hard to quickly estimate it and adjust,
  much less put it into some kind of optimization loop.
When tuning, we usually use proxies for this.

The simplest heuristic proxy statistic for tuning is the \emph{acceptance fraction}.
If you are accepting almost all proposed steps, your step sizes are too small on average.
If you are accepting almost none, your step sizes are too large.
The Goldilocks value is between a half and about a quarter,
  with an argument floating around that it should be 0.234 for best
  performance in high dimensional problems\note{%
    The argument and assumptions underlying the 0.234 fraction
    (and higher acceptance fractions at lower numbers of dimensions)
    are laid out in \citet{gelman234}.\label{note:234}}
  (though you could never tune it precisely enough to warrant that third digit of accuracy).
In the burn-in phase (discussed below in \sectionname~\ref{sec:initialization}) of an \MCMC\ run,
  it makes sense to track the acceptance ratio,
  and adjust the proposal distribution variance as you get acceptance ratios
  that are far from the Goldilocks ratio.
This process can be automated easily;
  such automation is part of many projects that use \MCMC.

A very common---%
  and very useful---%
  kind of proposal distribution is one that cycles through parameters,
  taking a random step in just one parameter at a time.\note{This has a lot to do with Gibbs sampling, which is discussed in \sectionname~\ref{sec:methods}.}
This kind of proposal distribution can be valuable,
  in part because it reduces the (almost impossible) $D$-dimensional tuning problem
  to $D$ one-dimensional problems:
In this form of proposal distribution,
  it is possible to track a separate acceptance fraction for every parameter.
Code can be built that uses the burn-in phase to tune all $D$ proposal variances
  such that each of them, individually,
  obtains acceptance at the same Goldilocks ratio.

One note to make here---%
  because it is relevant to tuning---%
  is that tuning can \emph{only} take place during the burn-in phase
  (that is, some part of your chain you will discard later);
  you cannot tune \emph{while} you run your final \MCMC\ run.
Why not?
Because tuning the proposal distribution based on the past history of the chain
  violates the ``Markov'' property
  that each step depends \emph{only} on the state at the previous step.
Violating the Markov property can be very bad:
  when you violate it,
  you lose all the provable properties of \MCMC\
  on which all our righteous power is based.

Another proxy for autocorrelation time useful for tuning is the
  Expected Squared Jump Distance.\note{See \citet{esjd}.}
This is the mean squared distance the walker moves, per step.
It is maximized when the acceptance ratio is reasonable
  and the step size is large;
  it is large when the exploration of the space is fast.
This proxy is easy to measure and use for tuning,
  and more directly related to autocorrelation time than the acceptance ratio.
We recommend using it for tuning, though we have never used it ourselves.
Like the acceptance fraction, it can also be used to tune
  in the case that the proposal distribution loops over parameters.
Once again, the user gets (in this case) $D$ one-dimensional tunings.

When sampling really gets very slow,
  there are various tricks and tips to work through.
The huge bag of possible tricks is so large,
  it goes way beyond the scope of this introductory \documentname.
In our experience, it is valuable to make friends with at least one
  statistician, one applied mathematician, and one computer scientist.
Between them, they ought to span the relevant literature.

\begin{problem}\label{prob:tuning}
Run the \MCMC\ sampling of \problemname~\ref{prob:twod} with the
covariant Gaussian density.
Give the proposal density $q(x'\given x)$ a diagonal variance
tensor that is $Q$ times the two-dimensional identity matrix.
Assess the acceptance fraction as a function of $Q$.
Find (very roughly) the value of $Q$ that gives an acceptance
fraction of about 0.25.
Don't try to optimize precisely; just evaluate the acceptance fraction
on a logarithmic grid of $Q$ with values of $Q$ separated by factors
of 2.
\end{problem}

\begin{problem}\label{prob:tuningtau}
Re-do \problemname~\ref{prob:tuning} but instead of trying to reach a
certain acceptance fraction, try to minimize the autocorrelation time.
You will need one of the autocorrelation-time estimators you might have
built in a previous \problemname.
(This, by the way, is the Right Thing To Do, but often expensive.)
What do you get as the best value of $Q$ in this case?
Again, just evaluate on a coarse logarithmic grid.
\end{problem}

\begin{figure}[!htbp]
\begin{center}
\includegraphics[width=0.48\textwidth]{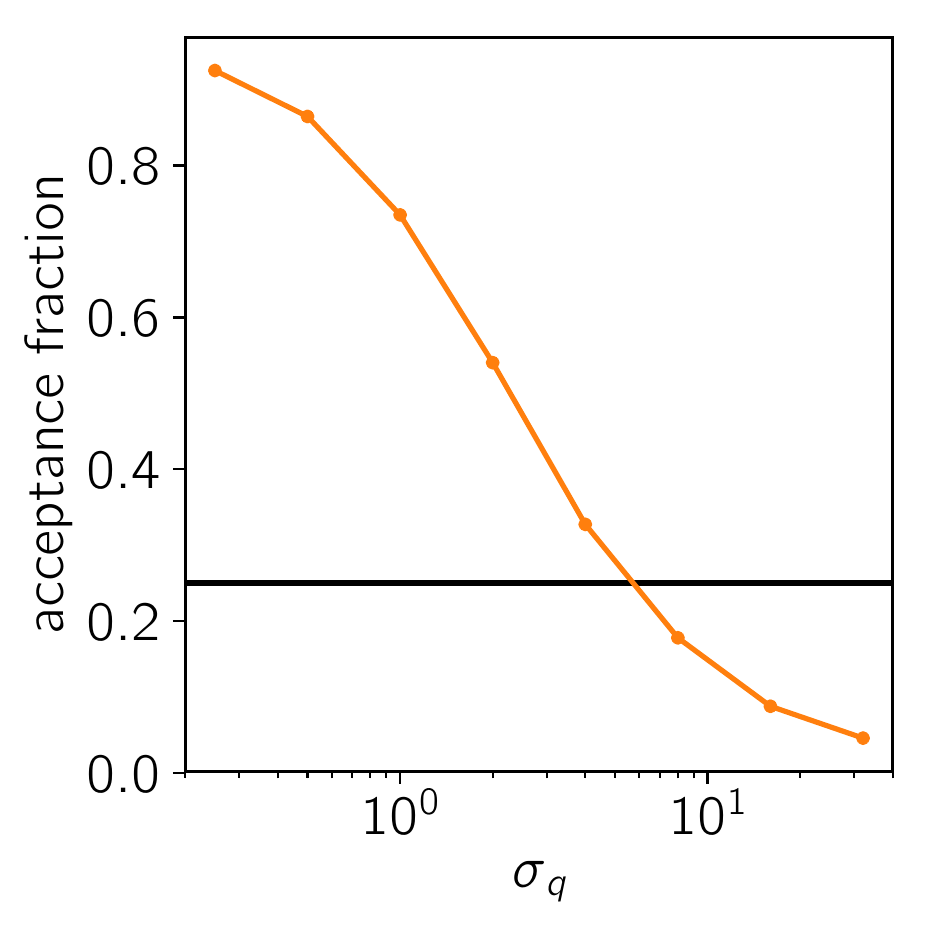}
\includegraphics[width=0.48\textwidth]{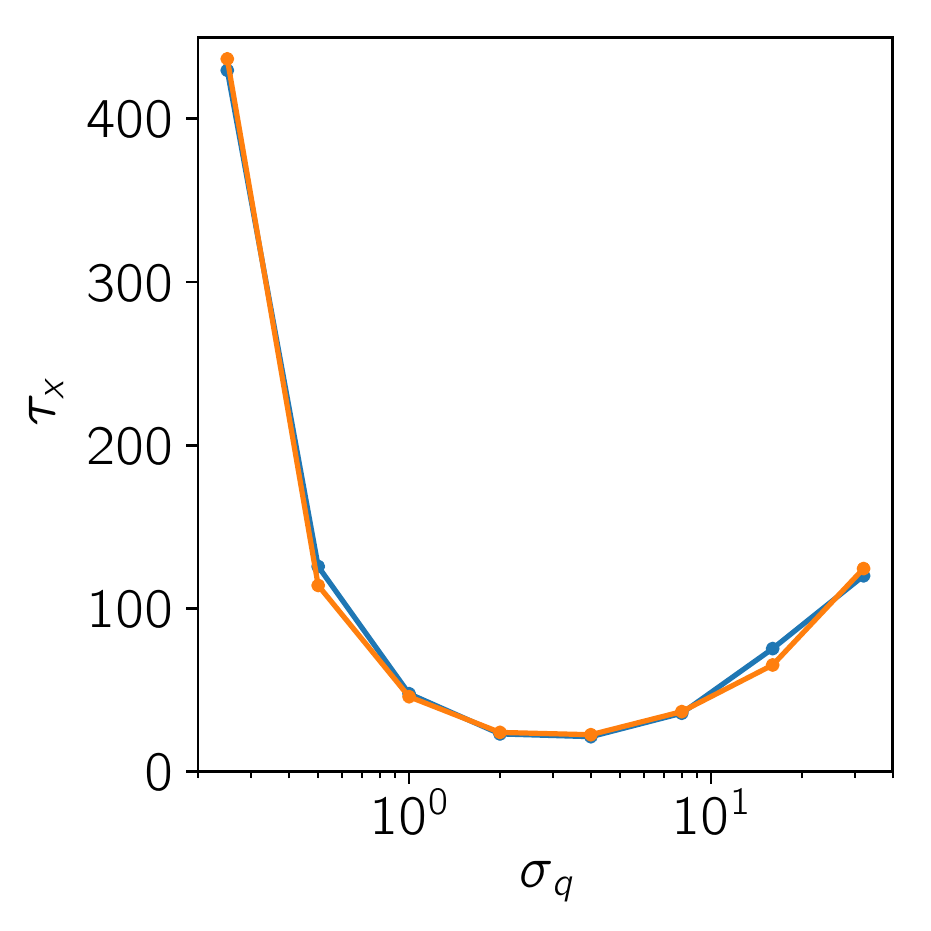}
\end{center}
\caption{Solutions to \problemname s~\ref{prob:tuning} and~\ref{prob:tuningtau}:
Two methods of tuning the M--H proposal parameters.
\emph{left:} The acceptance fraction as a function of proposal scale for the
distribution from \problemname~\ref{prob:twod}a.
\emph{right:} The integrated autocorrelation time for each parameter
(indicated by the different colors) as a function of proposal scale parameter.}
\label{fig:tuning}
\end{figure}

\begin{problem}
In \problemname~\ref{prob:tuning} you varied only the parameter $Q$, but
really there are three free parameters (two variances and a covariance).
If the problem was $D$-dimensional, how many tuning parameters would there
be, in principle?
\end{problem}

\begin{problem}\label{prob:rosenbrock}
The Rosenbrock density used as a demonstration case for many samplers (see, for example, \citealt{gw}).
Test your sampler on this density:
\begin{eqnarray}\label{eq:rosenbrock}
f(\theta_1,\,\theta_2) &=&
    \exp\left(-\frac{100\,(\theta_2-{\theta_1}^2)^2+(1-\theta_1)^2}{20}\right)
\quad.
\end{eqnarray}
Tune the Gaussian proposal distribution in your \MH\ \MCMC\ sampler to sample this density efficiently.
What autocorrelation time do you get?
Compare to what \project{emcee}\note{Available at \url{http://dfm.io/emcee}.} gets.
\end{problem}

\section{Initialization and burn-in}\label{sec:initialization}
\nopagebreak
Just as most (though not all) \MCMC\ methods require a choice of
proposal distribution, most (though not all) require a choice about
initialization:
The walker needs (or walkers need) to be started somewhere in the
parameter space.
In many use cases for \MCMC, the investigator wants to use \MCMC\ to obtain
uncertainty information about or propagate uncertainty into parameter
estimates.
In these cases, it makes sense to initialize the walker or walkers at
sensible parameter estimates, found by optimizing a likelihood or a
posterior pdf in advance of sampling.
In extremely high dimensions (large numbers of parameters), this can
be a bad idea, since the optimal parameters are not necessarily near
typical posterior samples\note{When the dimensionality of the
  parameter space gets very large, almost all samples will be
  \emph{very far} from the optimum of the posterior pdf.  One way to
  think about this is to think of a random Gaussian draw in $D$
  dimensions: A typical draw will be $\sqrt{D}$ standard deviations
  Euclidean distance from the maximum of the Gaussian.  That's a long
  way when $D=100$ or $1000$!  Another way to think about it is that
  there are a lot of ways to move \emph{away} from the point you care
  about, but very few ways to move close to it.  The upshot is that
  even with an \emph{enormous} sampling, if you are in large
  dimensions, not a single sample will be very near the optimum.\label{note:highD}}, but
in low dimensions (few to tens) this is often sensible.

Ideally, you will initialize the walker not at a completely irrelevant
point, nor at the optimum of the posterior pdf, but at a typical or
pretty good place in the posterior pdf.
That would minimize burn-in!
But, in general, being at the optimum or anywhere near a good place is
usually better than a mindless initialization.

As to burn-in:
If you have started your sampler in a non-typical place---or if you
are concerned that you \emph{might} have started your sampler in a
non-typical place---then you should discard the beginning of your
\MCMC\ run before you do your inferences.
This discarded part is called the ``burn-in''.
Some practitioners insist that burn-in cannot exist\note{\citet{geyer}
  takes this view.}: So long as the initial point is a conceivable
sample, you are fine!
Ensemble methods (like \project{emcee}\note{\citet{emcee}.}, and
mentioned below in \sectionname~\ref{sec:methods}) require a burn-in
phase and discard, because the whole point is that the ensemble must
grow (or shrink) to fill the posterior pdf volume, and you can't
initialize the ensemble sensibly if you don't know that volume
\foreign{a priori}.

If you suspect, or if it is even possible, that your problem is badly
multi-modal, then you will have to start at multiple points in
parameter space and compare the resulting chains.
By ``badly multi-modal'' we mean that there are peaks in the posterior
pdf that are connected by low-enough valleys that it is very unlikely
or takes a long time for a walker to traverse from one peak to the
next.
If this might be an issue, then the best diagnosis is to start many
chains in parallel with different initializations,
  and check that they lead to identical (or very
similar) posterior inferences (same pdf location and shape and
statistics).

Of course if you find out the worst---if you find out that different
initializations lead to different posterior inferences---you are in
trouble:
There is no trivial way to combine together the samplings you get of
different modes!
You can either wait a \emph{very long time} so that you see a single
walker in a single \MCMC\ run traverse from mode to mode enough times to
make a representative sampling (dozens or hundreds of times, ideally),
or you can use a high-end sampler (like nested sampling, see \sectionname~\ref{sec:methods}) that is
designed for such problems.
In principle there are methods that involve splitting the space into
two spaces, sampling them separately, and then combining the chains
according to their relative evidence afterwards.
That's a research project beyond the scope of this \documentname.

Of course there are some kinds of multi-modalities that might not be a
problem.
For example, you can have two solutions to a problem that differ only
in the \emph{labeling} of the components---say the Gaussians making up
a mixture of Gaussians---that make up the model.
One solution has the $K$ Gaussian components in one order, and the
other in another order, but they are identical Gaussians in all
other respects.
Because these two solutions make the same predictions for any data, and
only differ in their irrelevant, latent, internal ``naming'' of components,
they are not really different solutions.
This suggests that when you ask whether the outputs of two \MCMC\ chains
are consistent with one another, you should do so in the realm of the
parameters you \emph{care about}, not irrelevant parameters that do
not have impact on any present or future data.

But it is worth remembering that---just as \MCMC\ is not a good optimizer
(generically, samples will not lie close to the maximum of the posterior pdf)---%
it is also not a good \emph{search algorithm}.
There is no sense in which (standard) \MCMC\ methods are efficient at
searching, or engineered to search, all of parameter space.
If you really need to check all of parameter space, you should put
\MCMC\ aside, and do an exhaustive search (which will in general take an enormous
  amount of time).

\begin{problem}\label{prob:initialization}
Re-do \problemname~\ref{prob:MH} but with different starting
positions.
What happens as you make the starting position extremely far from the
origin?
What is the scaling:  As you move the initialization further away,
how much longer does it take for the sampler to reach reasonable
density?
\end{problem}

\begin{problem}\label{prob:initialization2}
Check the scaling you found in \problemname~\ref{prob:initialization}
with a higher-dimensional Gaussian (try, for example, a 10-d Gaussian).
The same or worse?
\end{problem}

\begin{problem}\label{prob:optimization}
Import (or write) an optimizer, and upgrade the code you wrote for
\problemname~\ref{prob:initialization} to begin by optimizing $\ln
p(x)$ and only then start the \MCMC\ sampler from that optimum.
Use a sensible optimizer.
Compare the scaling you found in
\problemname~\ref{prob:initialization} to the same scaling for the
optimizer.
To make this test fair, don't use the awesome math you know about
Gaussians to help you here; pretend that $p(x)$ is an unknown function
with unknown derivatives.
\end{problem}

\section{Results, error bars, and figures}\label{sec:results}
\nopagebreak
For a committed probabilistic scientist---frequentist or
Bayesian---there is no single ``answer'', there are only probability density
functions (pdfs).  In a paper or in an email or in a conversation
we might say what ``the answer is'', but even if we say it with an error
bar, we have departed the probabilistic program.  There are many
principled ways to depart the program; in another forum, we hope to say more
about the economic explanation of how we can make hard decisions in
the context of probabilistic reasoning\note{The idea is that you
  can make principled decisions by optimizing the expected utility under
  the posterior, given the data.  This is a great idea!  Of course we also
  have very deep, fundamental reasons that you \emph{can't know your utility
    precisely}.  The big issue is that your only sensible utility involves
  an integral out to the ``long term'', and the long term (by definition)
  includes outcomes that are outside your present-day quantitative model.}
But without going into that economic model, it is fair to say that a
substantial problem with being a committed probabilist is that when we
\emph{publish}, we are not permitted (not now, at least) to publish a
\emph{probability distribution over publications}; we have to publish
\emph{one single, deterministic text}; we have to make a decision about what
to write in the title, the abstract, the tables, figures, and results
section.  Given a \MCMC\ sampling of the posterior pdf, what do we
report as our \emph{results}?

One thing we can keep in mind to guide us in this is what we said above
  (in \sectionname~\ref{sec:sampling}),
  which is that samplings are good for doing \emph{integrals}.
We should endeavor to use as our ``results'' outputs from the \MCMC\
  that are based on integrals computed with the sampling.
This includes expectations, medians, quantiles, one-dimensional histograms,
  and multi-dimensional histograms.
This does \emph{not} include the ``best'' sample or a mode or optimum.
The latter things are not necessarily illegitimate outputs of the \MCMC,
  but they do not make best use of the fact that the sampling is a tool for integration,
  and we do not recommend them.

Imagine that you are in the simplest case:
You have a model with a small number of parameters (say three-ish),
  and only one of them is of great interest.
What are your options for the ``measurement'' of this parameter?
The only simple integral-based options are the posterior mean
  or posterior median value for the parameter.

For example,
  imagine that by \MCMC\ you have generated
  $K$ samples $\pars_k$ of a parameter vector $\pars$.
Now you have a scalar function $g(\pars)$ which takes the parameter vector
  and returns a scalar value%
  \note{For the median-of-sampling value, $g(\pars)$ needs to return a scalar,
  but technically, for the mean-of-sampling, $g(\pars)$ can be a vector or something high-dimensional.}
  which could be as simple as a single component of the parameter vector
  (one parameter from the list),
  or something more complex.
The mean-of-sampling value $\mean{q}$ is just
\begin{eqnarray}
\mean{q} &\leftarrow& \frac{1}{K}\,\sum_{k=1}^K g(\pars_k)
\quad ,
\end{eqnarray}
  and the median-of-sampling value is just the $[K/2]$th value of $g(\pars_k)$
  when the $g(\pars_k)$ have been ordered (sorted) from lowest to highest.
These are both produced by integrals;
  the first is the value returned by the expectation integral estimate,
  the second is the value past which half of the integrated pdf lies.

What are your best options for the ``uncertainty'' or ``limits'' on this parameter?
The only simple integral-based options are either variances or else posterior quantiles.
We usually use quantiles.
That is, the ``one-sigma'' error bar can be taken to be the half-size of the central (or smallest) interval of the parameter that contains 68 percent of the posterior
  samples.
In our own work, We usually create a 68-percent interval or region by
  excluding the top and bottom 16 percent of the
  posterior samples.  Some more aggressive investigators find the
  smallest interval that contains 68 (or 95) percent of the samples.
  Both are legitimate, and they are nearly identical for distributions
  that are nearly symmetric around the mean.
The ``two-sigma'' error bar would be the same but for the 95-percent interval
  (excluding the top and bottom 2.5 percent).
Again, these limits can be estimated%
  \note{One amusing thing about samplings,
  which are so beloved of us probabilistic (Bayesian) reasoners,
  is that anything we \emph{do} with a sampling,
  like estimate an integral or a quantile,
  is just an old-school frequentist estimator.
  A frequentist estimator of a Bayesian quantity, to be sure, but a frequentist estimator nonetheless.}
  by ordering the $g(\pars_k)$ values,
  and finding the $\pars$ values such that some fraction (0.68 or 0.95)
  of the samples lie between them.
Again, to form these limits, $g(\pars)$ must be a scalar function (so sorting is well defined).
Because they are so afraid of being confused with frequentists,
Bayesians often call these regions ``credible'' rather than
``confidence''\note{We don't like this ``credible'' terminology,
  though we sometimes adopt it to avoid confusion.}

One amusing thing is that if you choose the 68-percent interval sensibly,
  but use as the ``measurement'' the posterior mean,
  you can get pathological situations (from large skewness)
  in which the posterior mean measurement
  is actually \emph{outside} the one-sigma confidence interval.
For this reason (and others), we usually recommend as a default behavior%
  ---in the one-dimensional case---%
  to choose the \emph{median} of sampling as the measurement value,
  the 16-percent quantile as the lower one-sigma error bar,
  and the 84-percent quantile as the upper one-sigma error bar.
This has pathologies in higher dimensions (as we are about to see),
  but is pretty safe for one-dimensional answers.

The conservative scientist would show not just 68-percent error bars,
  but also 95-percent
  (to help readers visualize the skew of the posterior pdf).
The very conservative scientist would also show a histogram
  of the posterior samples,
  with the relevant quantiles (2.5, 16, 50, 84, and 97.5~percent) indicated.

As we mentioned above, scientists often like to report the ``best sample'';
  that is, the sample with the highest posterior pdf or $f(\pars)$ value.
This is not usually a good idea.
For one, \MCMC\ is a sampling algorithm, not an optimization algorithm:
  there are no guarantees that it will find the optimum of the posterior pdf in reasonable time.%
  \note{The sampler samples the pdf fairly.
    The tallest peak in the posterior is not guaranteed to be---%
    and in general is not---%
    also the peak with the greatest posterior ``mass''.
    That is, the samples in a fair sampling will not necessarily come
    predominantly from the \emph{tallest} peak in the posterior pdf.
    They will come from the peak with the greatest total integral
    of the likelihood over the prior.
    These issues might sound ``academic'',
    but when the number of parameters gets large (greater than, say, 10),
    many normal intuitions one might have (if any) about what is reasonably likely in a pdf
    are regularly violated. See also \notename~\ref{note:highD}.}
For two, the closeness of the best sample to the posterior mode is a strong function of sample size,
  and with very bad scaling.%
  \note{If you have a $K$-point sampling in a $D$-dimensional parameter space,
    there is some best sample $\pars_k^{\best}$.  Now imagine that you want to
    find a point $\pars^{\better}$
    that is ten times closer to the true optimum.
    In the limit that $K$ is large you have to take (on average) a \emph{factor} of
    $10^D$ more samples!
    Any reasonable optimizer is far, far better than this (in terms of compute time).}
For three, if you just want an optimum, use an optimizer!
That is, we \emph{strongly advise against} reporting,
  as ``the'' measurement or ``the result'',
  the parameter value or values for the best sample.
Giving some posterior samples is a good idea (we return to this below).

If you feel drawn to give the best sample, instead give the parameter
values found by optimizing the posterior pdf, starting at the best
sample as an initialization (and sacrifice your probabilism\note{We
  choose not to count the number of papers out there that claim to be
  ``Bayesian'' but then deliver the optimum of a posterior PDF.  That
  sacrifices probabilism, but also all the useful things you get from
  delivering a posterior, like protection from over-fitting, and
  non-approximate uncertainty propagation.  You certainly don't need
  to be a Bayesian---you can do most of the same science as a
  frequentist---but if you are using \MCMC\ you probably are doing so
  because you want the benefits of Bayesianism.}).
This optimal-posterior parameter value is called the
  ``maximum \foreign{a posteriori}'' or ``\acronym{MAP}'' value for the parameters:
  It is like a maximum-likelihood value but regularized by the prior.
It is an estimator with some good (and some bad) properties, but it is not the point of \MCMC,
  and therefore outside the scope of this \documentname.

The parameter space $\pars$ is usually multi-dimensional, and usually it is more
than one of those parameters that is of interest.
In this case, the mean or median of sampling can be produced for each
dimension.
Because the sampling projected onto any dimension is a marginalization
of the posterior pdf, any such one-dimensional mean or median is the
mean or median of the marginalized posterior pdf.

There is one oddity to note here, as it catches many investigators by
surprise:
Even if the model is good and the posterior pdf is unimodal and well-behaved,
the median (or even mean) of the posterior pdf for each of parameters,
when taken together as a ``best-fit'' parameter vector $\pars^\ast$,
will not itself necessarily be a good fit to the data!
If there is substantial ``curvature'' to the pdf in the parameter
space, the mean or median of sampling does not necessarily lie at a
high-probability location of parameter space.
This may seem counterintuitive, but it is easy to see in the case of a
``banana-shaped'' posterior pdf.
This all relates to the fact that \MCMC\ is not an optimizer, it is a
sampler.
It is important, when reporting the output of an \MCMC\ run by giving
means or medians of the posterior pdf to remind the reader or user of
that output that it does not necessarily represent (collectively) a
good fit to the data; these outputs only give information that is
useful one parameter at a time.

In principle the only output from the sampling that safely gives both
probabilistic information about the result of the inference and also
good-fitting models is a few---randomly chosen---example samples.
We strongly recommend this\note{And we do it ourselves.  For examples,
  you can look at \citet{cometholmes} and \citet{exopop}.}; in
particular this is a better thing to do than to give the ``best
sample'', which isn't even guaranteed to be near the bulk of the
posterior pdf or the samples therefrom.

If you \emph{do} return posterior samples to your audience, how many
should you return?
There are only heuristic answers to this, but a guideline is that, if
you only care about a single (one-dimensional) parameter, you don't
need many samples; a dozen suffices to give you a reasonable posterior
pdf mean and variance.\note{This point is made well by \citet{mackay}.}
If you care about $K$ parameters or dimensions of the parameter space,
in general the need for samples grows exponentially (or perhaps even
factorially!) with the number of dimensions.
So if you have a ten-dimensional space, expect to be publishing your
samples in an electronic table on the web!

If you \emph{do} publish sufficient numbers of posterior samples, you
might find users who want to make serious use of them.\note{We have
  been pushing this approach in our own work; our hierarchical
  inferences \citep{eccentricity, exopop} make use of importance
  samplings, starting at posterior samplings from interim prior pdfs.
  We have used this technology for computational convenience, but it
  is a framework for building subsequent analyses on the posterior
  samplings provided by other investigators.  One key---if you want
  your posterior samplings to be useful---is that you must release
  (along with your posterior sampling), the value of the prior pdf at
  the locations of the samples.  That is, you must ``decorate'' the
  samples (augment them) with prior pdf values or calls, or else
  release an executable version of your prior pdf, into which all your
  samples can be put as inputs.  In \emph{principle} it is enough to
  release a \emph{description} of your prior pdf, but in practice it
  is rarely correct or specific enough for another investigator to
  duplicate exactly.  So just augment your samples!\label{note:importance}}

Of course figures are always better than tables, at least in the
printed form of scientific publications.
There are several figures that are useful and should be made every time you
run \MCMC\ and we'll describe a few here.

\begin{itemize}
\item \emph{Trace plots}~---~The first set of plots are the ``trace
plots''~--~the parameter values as a function of step number.
These plots can be used to select burn-in lengths, indicate problems with the
model or sampler, and qualitatively judge convergence.
This being said, it is important to remember that \emph{good looking trace
plots do not guarantee converged sampling} and, even if the traces look okay,
the convergence diagnostics discussed in \sectionname~\ref{sec:convergence}
might identify problems.

\item \emph{Posterior predictive plots}~---~Another useful plot is the
predictive distribution for the model in data space.
For this plot, you take some $K$ random samples from your chain, plot the
prediction that each sample makes for the data and over-plot the observed
data.
This plot gives a qualitative sense of how well the model fits the data and it
can identify problems with sampling or convergence.

\item \emph{Corner plots or scatterplot matrices}~---~In a $D$-dimensional
parameter space, we recommend plotting all $D$ one-dimensional sample
histograms and all $D$-choose-2 two-dimensional histograms (scatter plots) to
show the low-level covariances and non-linearities.
If you are clever, all these plots can be arranged into a lovely and
informative triangle.\note{Our favorite package these days is our own
\project{corner.py}.  Find it on the internets.} These two-dimensional plots
do not in any sense contain all the information in the sampling but they are
remarkable for locating expected and unexpected parameter relationships, and
often invaluable for suggesting re-parameterizations and transformations that
simplify your problem.
\end{itemize}

\begin{problem}\label{prob:rosenbrock2}
Execute \problemname~\ref{prob:rosenbrock}, or---if you are lazy---%
just install and use \project{emcee} to do the hard work.
Now plot the $x$ and $y$ histograms of a 10,000-point sampling of this
distribution (you might have to sample more than 10,000 and thin the
chain), and also plot the two-dimensional scatter plot of $x,y$
samples.
Overplot on all three plots an indicator of the means and medians of
the samples along the $x$ and $y$ directions.
Overplot on all three plots the (above) recommended quantiles of the
samples.
Comment on the results.
\end{problem}

\begin{figure}[!htbp]
\begin{center}
\includegraphics[width=0.7\textwidth]{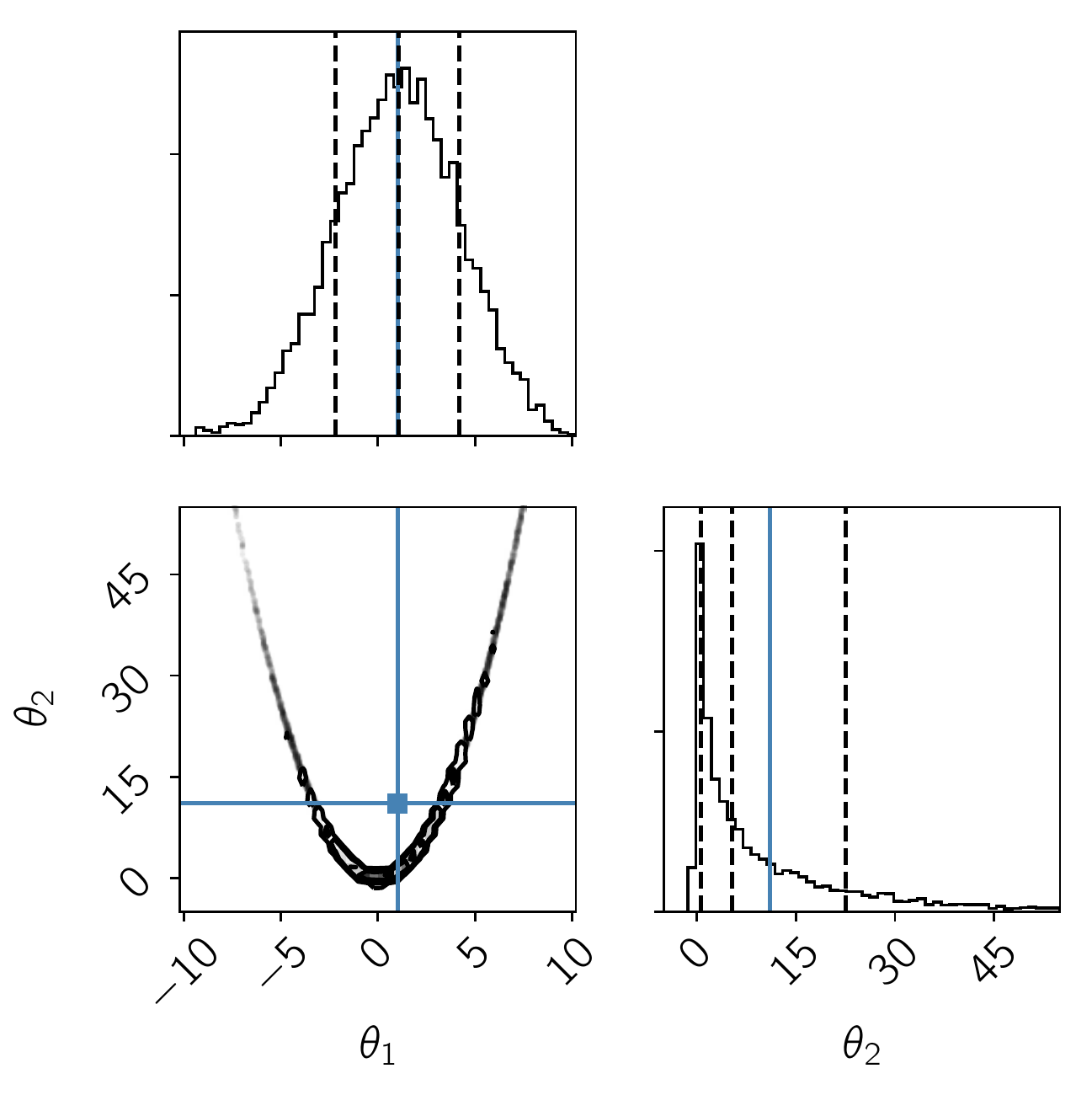}
\end{center}
\caption{Solution to \problemname~\ref{prob:rosenbrock2}:
\MCMC\ samples from the Rosenbrock density
(\equationname~\ref{eq:rosenbrock}) with the mean shown with blue lines and the
0.16-, 0.5-, and 0.84th quantiles shown as vertical dashed lines.
    \label{fig:rosenbrock2}}
\end{figure}

\section{Troubleshooting and advice}\label{sec:trouble}
\nopagebreak
When issues arise with \MCMC\ sampling,
  they are sometimes difficult to diagnose:
Is it the \MCMC\ code, the priors, the likelihood function,
  the initialization, the tuning, or something else?

\paragraph{Functional testing}
In addition to a full set of unit tests for every part of your code
  (yeah \emph{right}\note{If you aren't unit testing, you are probably making some very big mistakes. It makes sense to put in unit tests for every part of your code, especially the parts of your code that are relied upon for correctness in multiple locations, like your \MCMC\ sampler.}),
  we recommend building some end-to-end functional tests that permit you to perform \MCMC\
  runs with output that must meet certain expectations.
The simplest kind of functional test is to sample a known distribution,
  and check that the sampling has the moments you expect (to within tolerances).
For example, you can sample a $D$-dimensional Gaussian distribution with known non-trivial covariance,
  and check that the empirical covariance comes out as expected.
This is an easy test to fail, so a success builds confidence in any \MCMC\ code.

Another important functional test is to \emph{sample the prior pdf}.
This tests the sampler,
  tests that your priors really are what you think they are
  (and there are so many ways for them \emph{not} to be\note{It is
    easy to think you have flat priors when in practice your sampler
    has flat priors in the logarithms of the parameters or inverses of
    the parameters.  Also, it is easy to have improper priors when you
    think they are proper, because limits are not working, or you were
    wrong about the convergence of some integral.  Sometimes the prior pdf in practice has
    different edges than you think because there is some censoring
    happening that you haven't considered.  And then, of course, it is
    possible for your sampler itself to be buggy.}), and tests that
  your prior pdfs are proper (functions that can be normalized).
This test is most effective as a true end-to-end test of your model
  if you structure your code such that the log probability function \code{ln_f}
  input to the \MCMC\ scheme
  is set up as a sum of a log-likelihood function \code{ln_likelihood}
  and a log-prior function \code{ln_prior}.
The sampler can be switched to sampling the prior by a simple replacement
  of the \code{ln_likelihood} function
  with a trivial function that always returns zero.

\MCMC\ should run well in this flat-likelihood case (priors are usually easy to sample)
  and the output sample histograms should look like what you expect
  given the prior pdfs you coded.
If the walkers run off to positive or negative infinity,
  and nothing seems to converge,
  your priors are probably not proper.
If the priors don't look as you expect,
  you either have a bug in your code,
  or else a think-o about how your priors ought to look.
Either way, diagnosis is in order.

\paragraph{Likelihood issues}
Sometimes there can be problems with the likelihood function itself.
For example,
  when you re-call the likelihood function with the same parameters,
  do you get \emph{exactly} the same answer?
If you don't, your likelihood function effectively depends on some random numbers;
  maybe it includes within it an integral performed by Monte Carlo method?
In general, you want to write your likelihood function such that,
  even if it includes a numerical or stochastic integral,
  it is designed such that it produces \emph{precisely} the same
  value when it returns to the same point in parameter space.
One way to make this happen is to choose the random numbers
  (those used for any internal integration inside the likelihood function)
  in advance and re-use them on every likelihood call.
This provides the same integral according to the same approximation,
  but both speeds up the code (no re-generation of random numbers)
  and also makes the likelihood function single valued.
Even better is to replace internal Monte Carlo integrals with
  deterministic numerical (or even better analytic) integrals.

If things are giving trouble and you suspect the likelihood function,
  another important test is to visualize slices through parameter space.
When you call the likelihood function on a grid in each parameter
  (with the others fixed, say, to reasonable values),
  you should see a likelihood value that is a smooth function of each parameter,
  or at least as smooth as you expect.
Things like numerical integration,
  truncated expansions,
  or adaptive approximations
  inside likelihood functions
  can make the function noisy or jagged at small scales in parameter space.
These problems hurt optimizers more than samplers (although Hamiltonian
  methods depend strongly on good derivatives, see \sectionname~\ref{sec:methods}),
  but they usually point to code issues.
In general, plotting the likelihood function as a function of parameters along
  slices in parameter space often reveals bugs or think-os.

\paragraph{Low acceptance fraction}
If you find that you are getting low acceptance fraction
  (in standard M-H \MCMC\ or other varieties for which there is such a concept)
  you can simply reduce the sizes of the steps you consider to increase the fraction that are accepted.
This means decreasing the variance of the proposal distribution,
  or whatever parameter or parameters control the width of that distribution.
If reducing the variance of the proposal distribution does \emph{not}
increase the acceptance fraction, then \emph{you have a bug}.
You must find that bug by auditing your code or else returning to the
functional tests listed above.

Similarly for high acceptance fraction:
If you are finding a high acceptance fraction, then increasing the
variance of the proposal distributon \emph{must} decrease the
acceptance fraction.
If it does \emph{not}, then \emph{you have a bug}.

\paragraph{Sticky chains}
If you are concerned about acceptance fraction or convergence,
  it makes sense to plot parameter values as a function of ``step number'' or iteration number.
That is, plot the ordered chain in each parameter dimension.
These plots will have long horizontal patches if the chain is getting stuck;
  a converged chain will show the walker traversing the parameter space in every dimension fully many times
  over the length of the run.
Not a few times; many times.
If you are using an ensemble sampler\note{Such as \project{emcee} \citep{emcee};
  one nice thing about the ensemble sampler is that it gives you great diagnostic information automatically.}
  make plots showing all $M$ walkers, each a different color.
A converged run for an ensemble sampler will show every single walker traversing the parameter space
  in every dimension many times.

\paragraph{Initialization-dependence}
If you initialize your \MCMC\ runs in different places in parameter space,
  do you get the same (or very similar) final parameter samplings,
  in mean and variance?
If you don't,
  then your posterior pdf is badly multi-modal;
  trivial \MCMC\ methods probably won't fix your problem.
Not to put too fine a point on it:
If your results are initialization-dependent,
  then your \MCMC\ runs \emph{are not converged}.

You either have to go to a method that tries to explore the posterior
  more liberally, such as nested sampling or simulated tempering,
  or else split your ``model'' up into several sub-models,
  each of which contains different posterior modes.
In general, this problem is hard to fix;
  as we mentioned above, it is provably impossible
  to explore all of parameter space if the posterior pdf is
  complex in morphology.
We don't have very useful advice here, except to
  spend time learning about the multiple modes in the posterior pdf
  and what each ``means'' or corresponds to,
  and to do your best to express in your results the multiple optima.
One of the miracles of \MCMC\ is that if you \emph{can} fairly sample
  as complex posterior pdf,
  the fraction of samples in each mode
  tends to the relative total integrated posterior probability inside each mode,
  as the sampling converges.

Ensemble samplers react to multiple modes
  by obtaining very low acceptance ratio\note{Although maybe this isn't
    fundamental; see \project{kombine} at \url{https://github.com/bfarr/kombine}.},
  because the samplers in the different modes are not easily able to ``help one another''
  move efficiently.
Again, if the acceptance ratio is low and there appear to be multiple modes,
  it is likely that the sampling will also be initialization dependent,
  and it is very unlikely that your sampling is going to be converged.

\paragraph{Bad initialization}
If you initialize your sampler in a disallowed region of parameter
space---that is, a part of the space with $f(\pars)=0$ (so your
\code{ln_f()} function returns \code{-Inf})---it might have a very
hard time random-walking out of that location.
At initialization time, it is important to test that the
initialization is permitted.
For ensemble methods, it is important to test that \emph{all} the
walkers in the ensemble are initialized to permitted values.

\paragraph{Parameterization problems}
It helps to think carefully about how to parameterize.
For example, if there is an angle $\phi$ in your problem,
  and your prior requires it to be between $0$ and $2\pi$,
  and the posterior mode is very near $0$,
  a little bit of probability leaking below $0$ plus the prior cutoffs plus mod-$2\pi$ symmetry
  makes an intrinsically unimodal problem multimodal!
In these cases, we usually advise reparameterization from a (say) amplitude $A$ and angle $\phi$
  to two vector components $a\equiv A\,\cos\phi$ and $b\equiv A\,\sin\phi$.
Of course this re-parameterization changes significantly the \emph{form} of the prior pdf
  (the transformation brings in a Jacobian);
  it has to be adjusted with care.

To be explicit, imagine that you have parameters $A$ and $\phi$ and
priors $p(A)$ and $p(\phi)$.
Now imagine that instead (and very sensibly) you want to sample in
parameters $x\equiv A\,\cos(\phi)$ and $y\equiv A\,\sin(\phi)$.
What is the proper prior to apply to $x, y$?
It is not just $p(A)\,p(\phi)$ with $A=\sqrt{x^2 + y^2}$ and $\phi=\arctan(y,x)$!
That's close, though! There is a correction factor which is the
absolute value of the determinant of the Jacobian matrix
$||{\dd(A,\phi)}/{\dd(x,y)}||$.
That is, it is the determinant of the full derivative of one set of
parameters with respect to the other.
Keep your units straight!\note{See \citet{hoggcalculus} for advice on
  how to check and test these transformations using dimensional
  analysis.}
In this particular case, the Jacobian is equal to $1/A$, such that
\begin{eqnarray}
  p(x,y) &=& \frac{1}{A}\,p(A)\,p(\phi)
  \quad ,
\end{eqnarray}
where $A=\sqrt{x^2 + y^2}$ and $\phi=\arctan(y,x)$.\note{Exercise to
  the reader: Do the same but with $x\equiv \sqrt{A}\,\cos(\phi)$ and
  $y\equiv\sqrt{A}\,\sin(\phi)$.}

For another example, sometimes there is almost an exact degeneracy between two parameters,
  say $a$ and $b$.  Then it makes sense to switch to parameters $a+b$ and $a-b$,
  or some other linear combinations where correlations or near-degeneracies are likely to be reduced.
Again, such transformations require also prior transformations, which must be made with care.
There are some affine invariant samplers\note{Note how much we like to
  cite \project{emcee} \citep{emcee}!} that are invariant to such
transformations, but if you aren't using such a sampler, it is worth
getting out ahead of these problems.
They can be found by performing an exploratory sampling,
  making a triangle plot as described above,
  and looking for narrow diagonal lines in the two-dimentional scatter plots.

For another example, sometimes some parameters are continuous, and some
are integer (or discrete).
In this case, it is sufficient, in the context of M-H \MCMC, to make
custom proposal distributions for the integer parameters that make sure
only integers are proposed (provided care is taken to make sure that detailed
balance is preserved).
An example is to have a proposal that has some probability of
incrementing the parameter up, and an equal (that is, the same) probability
of incrementing it down.
Some samplers implicitly or explicitly assume that all parameters are
continuous\note{Once again, \citet{emcee}.}; these should be avoided
when some parameters are integer.
If you \emph{really} want to use such samplers when some parameters
are discrete, you have to put a \code{round()} or \code{int()}
operation into the likelihood and prior functions.
This makes for a stepwise-constant posterior pdf. It is a hack, but it
works in most circumstances.

For yet another example, there might be constraints such that all
acceptable models lie on a non-trivial, non-linear subspace of your
parameter space.
Here almost no arbitrary steps in parameter space could lead to an
acceptance, since if you move arbitrarily, you are very unlikely to
hit the subspace!
Your best move here is to re-parameterize so there are only models on
that subspace.
We're not claiming that that re-parameterization is always easy to
find; the mathematics of Lagrange multipliers can be useful here.

\paragraph{Model checking}
\emph{All models are wrong, but some are useful.}
That's not a quote from Box, but it's close.\note{The real quotation
  is a parenthetical sentence ``Remember that all models are wrong;
  the practical question is how wrong do they have to be to not be
  useful.'' (\citealt{box}, p.~74).  It is amazing how rarely this is
  ever quoted correctly; or maybe there is another source for this
  quotation?}
The point (for our purposes) is that if you check your model hard
enough---that is, you take enough data---you will certainly rule it
out.\note{Physicists sometimes forget this, and believe that various
  physical theories, such as quantum electrodynamics or the
  cold-dark-matter cosmological model, are strictly True with a capital ``T''.  First of
  all, there is no way to know that for sure, and second of all, even
  if they \emph{are} true, they don't precisely explain any
  observation, which is also affected by various kinds of auxilliary
  effects and noises for which there is (and can be, for deep reasons,
  to be discussed in other documents in this series) no extremely precise model.  So in
  detail, any model of any specific observation must be wrong, even in
  the (exceedingly unlikely) event that the fundamental model is
  correct.\label{note:truth}}
``Model checking'' is an enormous subject that goes way beyond the
scope of this \documentname, for which reason we won't say much about
it here except to make two points:

The first---and most important---point about model checking is that
you always learn a huge amount by checking the model.
What this entails is looking at the distribution of residuals or the
quality of the predictions of the model when compared to the data.
There are methods (like the chi-squared statistic and the Bayesian
evidence integral) that check the absolute quality of the model in a
scalar way.
These are usually uninformative, because (as we noted), all models are
wrong, and eventually the data will be good enough to return a bad
statistic here.\note{Sometimes Bayesians like to complain that the
  chi-squared statistic is a measure of the size of your data!  That's
  true when your model is wrong (that is, always); see \notename~\ref{note:truth}.}

Much more informative are methods (like plotting residuals of the data
away from the model in the space of the data) that look at what parts
of the data the model ``explains well'' and what parts of the data the
model ``explains badly''.
These kinds of experiments usually reveal inadequacies of the model in
a rich way, or confirm that certain assumptions are bad.
Cross-validation\note{Cross-validation is an incredibly valuable set
  of techniques that all astronomers should know.
  it is not clear what to cite here, but one possibility is \citet{crossval}.}
is a framework for constructing
valuable tests of this form, as is posterior predictive checks\note{%
  The posterior predictive check is one of the Bayesian equivalents of
  a goodness-of-fit test \citep{ppcheck}.}
Importantly, you want to find ways to compare your model to your data
\emph{in the space of the data}.
Don't just look at the parameter space of your model!
It is the data that really exist, especially when all models are wrong!

There are model checking methods (like varying the prior and
re-running \MCMC) that test the sensitivity of results to assumptions.
Not only are these very valuable for testing the model, they are
essential for writing any paper about your results: The reader always
wants to know the sensitivity of results to assumptions.\note{And be
  sure not to forget that there are (at least) as many assumptions
  encoded in your likelihood function as there are in your prior
  pdfs, probably far more. No data analysis is protected from \emph{subjectivity}.}
Once again, it is important to find ways to display the results of
such tests in the space of the data.
All we will say here about these kinds of rich model tests is that you
should do them, and report the results.

The second---and less important---point about model checking is that
the ur-Bayesian thing to be doing is computing a Bayesian evidence
integral.
However, it is usually inadvisable:
The integral is expensive to compute.
Most \MCMC\ methods are cleverly designed to \emph{avoid} computing this
integral (as we note above in \sectionname~\ref{sec:when}); so it usually adds
expense to the project.
At the end of the day, it returns only a single, scalar number, with
no guidance about how to use it (and indeed there are only heuristics
to reach for).
If your model is wrong, the richer tests of visualization of residuals
or sensitivity to assumptions are much more likely to lead to insight
about what changes to make or what new directions to explore.

\section{More sophisticated sampling methods}\label{sec:methods}
\nopagebreak
As we have discussed above (\sectionname~\ref{sec:MH}),
\MCMC\ methods are proven to be correct ``in the limit''; that is, when
an infinite number of samples have been taken.
When the function $f(\pars)$ has challenging properties, the results
can approach this limit very slowly (or not at all\note{You can get
  essentially infinite autocorrelation times if the function has
  isolated regions of finite density separated by large seas of zero
  or near-zero density.  One disturbing thing about these situations
  is that any empirical measure of autocorrelation time will give a
  finite answer, but the density won't be properly sampled, ever. This
  problem is related to a problem that any simple empirical measure of
  the autocorrelation time will be an under-estimate. In
  principle you can find these badly multimodal cases
  with multiple re-starts of the chain
  and (things like) the Gelman--Rubin diagnostic \citep{gelmanrubin}.}).
For this reason, in the finite duration of a scientific project, most
users of \MCMC\ have one of two (related) problems:
Either \textsl{(1)}~it is taking too long---it is taking too many steps or
executing too many calls of the function $f(\pars)$ to get independent
samples---or \textsl{(2)}~it is not
exploring the full parameter space---there are local optima (or bad
local geometry around those optima) ``trapping'' the algorithm.

There is no general, problem-independent solution for either of these
problems.
Indeed, there are many different strategies that work well or badly
for different functions $f(\pars)$.
In this \sectionname, we will try to guide the reader towards methods
that might help in different circumstances.
We will not give a full description of all the \MCMC\ algorithms out
there, but just provide some field notes and pointers to references.
New \MCMC\ methods are being developed all the time, so a reader with a
serious problem would do well to consult with the applied-mathematics
or statistics literature.\note{Your best interface to these
  literatures is often a colleague in another department, like applied
  mathematics, statistics, or computer science.  Break down those
  walls!}
The different methods to which we refer here are (in general) very
different in their difficulty of implementation.
We won't help with that either, except to say that there are more and
more open-source or distributed packages every year.
We will try to name some current implementations in the notes, but any
list compiled today will be incomplete tomorrow.\note{One check-point
  in this literature is the compilation volume by
  \citet{mcmchandbook}, but much has been developed and changed even
  since that was written.}
The spaces of \MCMC\ methods and \MCMC\ method implementations are both
growing rapidly, in part because \MCMC\ has become such a core
technology in the empirical sciences.

All we are going to do here is give a cursory mention to some methods
we know about, with some keywords that can be used to search more
deeply.
There is no sense in which this list is comprehensive; these are just
methods (or method classes) that are likely to be useful to projects similar
to projects we ourselves have executed.

\paragraph{Ensembles}
As we discussed (\sectionname~\ref{sec:tuning}), one of the critical challenges in the use of
\MCMC\ is tuning the proposal pdf $q(\pars'\given\pars)$.
In problems in which all of the parameters (all of the components or entries of $\pars$) are somehow ``equivalent'',
or the parameters can be seen as components of a vector in a vector
space, there are \emph{ensemble methods} that make use of not just one
random-walker but instead many walkers to automatically generate a properly tuned
proposal distribution.
These methods make use of the distribution of a set of independent
walkers in the parameter space ($\pars$-space) to gauge the typical step-sizes and
directions at which new proposals should be made, obviating or greatly
reducing the tuning requirements.
This tuning improvement comes at the cost of a burn-in phase at the beginning
of any run---a phase that lasts a few autocorrelation times---in which the
ensemble expands or shrinks to fill the posterior volume.
We are co-authors on one of these methods,
\project{emcee}\note{\citet{emcee}.}, which is popular in the
physical sciences, but there are others\note{A few ensemble methods that
are being used in astronomy: Differential Evolution \MCMC\ (\citealt{demcmc,
run-dmc}) and \project{kombine} (Farr \& Farr, in preparation).}.
It is worthy of note that ensemble methods have particular
peculiarities about initialization and convergence diagnostics that
are worth understanding before use.\note{For example, we find
  empirically that it is better to initialize the ensemble in a ball
  that is smaller than the full posterior pdf width, not larger.  For
  another example, the ensemble produces a set of chains that can be
  used to compute the Gelman--Rubin diagnostic \citep{gelmanrubin},
  but this is not a conservative test if the initialization was made
  in a small ball.}
Because ensemble methods are updating independent \MCMC\ chains, they
are often easy to parallelize to make use of multiple cores.

\paragraph{Gibbs}
Neither \MH\ \MCMC\ nor most ensemble methods good at going to very
  large numbers of parameters.\footnote{See the excellent rant ``Ensemble Methods are Doomed to Fail in High Dimensions'' at \url{http://andrewgelman.com/2017/03/15/ensemble-methods-doomed-fail-high-dimensions/}.}
Even dozens of parameters can be enough to drive autocorrelation times
  extremely large.
With larger numbers of parameters, new methods need to be considered.
Gibbs sampling is worth considering when the number of parameters is
  large, and the different parameters have different ``scopes'' in the problem.

In many problems that have large numbers of parameters,
  the parameters---the components or entries of the
  blob $\pars$---are not equivalent at all.
Some are global parameters, which touch or have an effect on every
sub-part of the data blob $\data$, while others are local parameters,
which only touch a small subset of the data.
Or, in another use case, some parameters are linear parameters
that---at fixed values of the other parameters---will have a Gaussian
likelihood or posterior pdf.
In these problems, there are some directions in parameter space that
are hard to move in (they require complete re-calculation of the full
non-linear function $f(\pars)$, and other directions that are very
easy to move in (they require only partial re-calculations or the
sampling could even be analytic and exact).
For problems with these structures, Gibbs samplers are ideal, and
there are good implementations.\note{For example, \project{Stan}
(\url{http://mc-stan.org}).}

A trivial kind of Gibbs sampling was mentioned above (\sectionname~\ref{sec:MH}) in the
discussion of \MH\ \MCMC:
Even when running vanilla \MH\ \MCMC, it is often useful to update only
one parameter at a time; that is, do only axis-aligned moves in the
$\pars$-space, cycling through axes as you go from step to step.
This helps with tuning and diagnosis, but it also permits a good
implementation to capitalize on simpler calculations for the simpler
parameter directions.\note{%
  The point is that sometimes there are some parameters that only
  affect a very small number of data points, or parameters that change
  the model or the likelihood function and prior values in a very
  simple-to-compute way.  If such simple parameters exist, it makes
  sense to write your likelihood and prior code to capitalize on these
  simplicities.  This is not always trivial, because it may involve
  caching parts of the calculation and so on.  Details would be beyond
  our scope.}

Gibbs sampling can also be very good for exploiting multiprocessing:
Global parameter updates might require a serial re-calculation of the
full likelihood, but local parameter updates can be done in parallel,
with each local parameter working on its local bit of data all at the
same time.
This kind of structure is common in hierarchical inference, where
local parameters touch only individual objects in some population,
say, and global parameters are parameters \emph{of} that population as
a whole.
Parallelization is built in to some Gibbs implementations.\note{For
  example, \project{JAGS} (\url{http://mcmc-jags.sourceforge.net/}) automatically
  parallelizes Gibbs problems.}

\paragraph{Hamiltonian}
At time of writing, the premier approaches for problems with large numbers
  of parameters are Hamiltonian methods.\note{For a review of the theory
  behind these methods, see \cite{hamiltonian}.}
These methods perform very well at large dimensions, but require (for speed)
  analytic derivatives of the function $f(\pars)$ with respect to the parameters.

In many cases of interest, the function $f(\pars)$ can be analytically
differentiated, such that it is possible to quickly execute a gradient
$\dd f/\dd\pars$ evaluation.
Indeed, in the modern world of computing, there are even
auto-differentiation systems that deliver automatically code that
produces the gradient of any function you can write.\note{%
  The idea is that for (almost) any function you can write,
  a robot can write the function that is the derivative of that
  function, using simple differentiation rules and the magic of the
  chain rule. Differentiation is a typographic operation on computer
  functions!  The idea is \emph{not} to take numerical (small
  difference) derivatives of the code; there is almost no way that
  finite-difference differentiation could help your sampling in the
  long run.}
There is a class of \MCMC\ methods that make use of gradient information
to effectively inform the proposal distribution and thus speed sampling.

The reason these methods are ``Hamiltonian'' is that they use a kind of
  Hamiltonian dynamics to improve the sampling: They augment the position
  of the walker in parameter space with a fictitious momentum in
  parameter space, and use the derivatives in the accept-reject step.
The momentum helps the sampler move through the space on a more efficient
  variant of the random walk.
The ergodicity property of (relevant) dynamical systems ensure good sampling (in
  the infinite-time limit, of course); the connection to dynamics is that
  the negative logarithm of $f(\pars)$ becomes the potential for the
  dynamical system.
At the present day,
  these samplers are essentially the only good options when the number of
  parameters gets very large.

\paragraph{Importance}
Sometimes it is possible to sample efficiently from a distribution
that is close to the target function $f(\pars)$ you care about, even when the target $f(\pars)$ itself is
hard to sample efficiently with \MCMC.
An example is a problem that has a simple (Gaussian, say) likelihood
function times a non-trivial prior:
It is easy to sample from a Gaussian likelihood times a Gaussian
prior, even without \MCMC, but it might be hard to do the same sampling
under a more complex prior.
Importance sampling is a direct sampling method (not really an \MCMC\
method at all):
You take a sampling from the approximate function (which by assumption
was easy) and then you re-weight the sampling (or do a subsequent
rejection of samples) using the ratio of the true function to the
approximate function as a weight or probability.
We use this quite a bit in our hierarchical inferences\note{%
  For example, we use importance sampling to convert individual-planet
  exoplanet inferences (all performed with dumb priors) into a full-population
  consistent hierarchical model in \citet{eccentricity}. See also \notename~\ref{note:importance}.},
and in problems where the dimensionality is low.\note{See, for example, \citet{joker}.}

Two of the significant technical details for importance sampling are the
following:
The approximate function must have the same (or greater) support as
the true function, and the method degrades in efficiency as the two
functions diverge.
If the ratio of the true function to the approximate function is to be
used as a probability for rejection of samples, then the true function
must be greater than or equal to the approximate function everywhere!

\paragraph{Tempering}
A huge problem for \MCMC\ samplers---which make local moves based on
local positioning in the parameter space---is that there can be
multiple modes in the function $f(\pars)$, separated by valleys of
zero or near-zero density.
A vanilla \MCMC\ method cannot easily cross such valleys with local
moves.
The idea behind simulated tempering and related methods is to ``smooth
out'' the function $f(\pars)$ or reduce its dynamic range, to make the
peaks less tall and the valleys less deep, to permit the \MCMC\ random
walk to cross the valleys.
The standard method is to introduce a ``temperature'' variable and,
while sampling, take the likelihood function to the power of the
inverse temperature (or, if working in the logarithm, multiply the log
likelihood by the inverse temperature).
When the temperature is high, this reduces the influence of the
likelihood function relative to the prior pdf, and (assuming that the
prior pdf is easy to sample) makes movement in the parameter space
easier.
As with Hamiltonian methods, the sense in which this system makes use
  of a ``temperature''
  is that the logarithm of $f(\pars)$ can be related to a potential,
  in a physical analogy.
At each step, when the temperature has moved away from unity, the
distribution being sampled is not exactly $f(\pars)$, but the results
can be transformed into a fair sampling.

\paragraph{Nested}
Similar to tempering-like methods are a class of methods called ``nested''
samplers\note{For example, see \citet{skilling} and \citet{dnest}},
which also smoothly change the target distribution away from $f(\pars)$.
In the nested case, instead of increasing the temperature, the
samplings are of a censored version of the prior, censored by the
value of the likelihoood function.
When the likelihood censoring is strong, only the most high-likelihood
parts of the prior get sampled; when the likelihood censoring is weak,
almost all of the prior gets sampled.
The idea behind nested sampling---like tempering---is that it is
designed to be a good method for exploring the full parameter space or
searching for all of the modes of the posterior pdf.
It is also designed such that it produces not just a sampling but also
an estimate of the integral $Z$ of the likelihood times the prior
(the Bayes factor or fully marginalized likelihood).

\paragraph{Reversible-jump}
Finally, there are some extreme problems in which it is not just hard
to sample in the parameter space, but the parameter space itself is
not of fixed dimension.

For example, if you are modeling an astronomical image with a
collection of stars, and you don't know how many stars to use; in this
case the number of stars itself is a parameter, so the number of
parameters is itself a function of the parameters.\note{Our attempt to sample in this very problem is \citet{brewer}.}
In these cases you can only use certain kinds of samplers (though M--H
\MCMC\ is one you can choose), and, in addition, you need to make
special kinds of proposals that permit the system to jump from one
parameter space to another that has different dimensionality.
In these cases, the concept of detailed balance becomes non-trivial,
but there are criteria for creating reversible-jump proposals between
the parameter spaces.
These methods are best built and operated under the supervision of a
trained professional.

\acknowledgements
We would like to thank
          Alex Barnett (Flatiron),
          Jo Bovy (Toronto),
          Brendon Brewer (Auckland),
          Will Farr (Flatiron),
          Marla Geha (Yale),
          Jonathan Goodman (NYU),
          Fengji Hou (NYU),
          Dustin Lang (Toronto),
          Phil Marshall (KIPAC), and
          Iain Murray (Edinburgh)
        for valuable advice and comments over the years, and
          Alessandro Gentilini
        for a very close read.
        This project benefitted from various research grants, especially
          NSF grants AST-0908357, IIS-1124794, and AST-1517237,
          NASA grant NNX12AI50G,
          and the The Moore-Sloan Data Science Environment at NYU.
This work was performed in part under contract with the Jet Propulsion
Laboratory (JPL) funded by NASA through the Sagan Fellowship Program executed
by the NASA Exoplanet Science Institute.

\raggedright

\end{document}